\documentclass[12pt]{article}
\usepackage{amsmath}
\usepackage{graphicx,psfrag,epsf,color,subfigure}
\usepackage{enumerate}
\usepackage{natbib}

\usepackage[unicode=true,pdfusetitle,
   bookmarks=true, 
   bookmarksnumbered=false, 
   bookmarksopen=false,
   breaklinks=false, 
   pdfborder={0 0 1}, 
   backref=false, 
   colorlinks=false]{hyperref}

\usepackage{amssymb}
\usepackage{multirow}
\usepackage{array}
\usepackage{bm}
\usepackage{booktabs}
\usepackage{float}
\usepackage[font=small,labelfont=bf]{caption}
\setcitestyle{citesep={;}}
\usepackage[total={6.4in,8.6in}]{geometry}
\usepackage{algorithm2e} 

\newcommand{\Mean}{{\mathbb{E}}}

\newcommand\independent{\protect\mathpalette{\protect\independenT}{\perp}}
\def\independenT#1#2{\mathrel{\rlap{$#1#2$}\mkern2mu{#1#2}}}
\let\code=\texttt
\let\proglang=\textsf 

\usepackage{authblk}

\newtheorem{thm}{Theorem}[section]
\newtheorem{coro}{Corollary}[section]

\newtheorem{lemma}{Lemma}[section]

\newtheorem{remark}{Remark}[section]

\DeclareMathOperator*{\argmin}{arg\,min}
\DeclareMathOperator*{\argmax}{arg\,max}

\begin{document}

\def\spacingset#1{\renewcommand{\baselinestretch}%
{#1}\small\normalsize} \spacingset{1}
%
%
\title{\bf GEAR: On Optimal Decision Making with Auxiliary Data}
\author[1]{Hengrui Cai   \thanks{hcai5@ncsu.edu} } 
\author[1]{Rui Song\thanks{rsong@ncsu.edu}}
\author[1]{Wenbin Lu\thanks{wlu4@ncsu.edu}} 
\affil[1]{Department of Statistics, North Carolina State University} 
 \date{}
 \maketitle  

\baselineskip=21pt

\begin{abstract} 
Personalized optimal decision making, finding the optimal decision rule (ODR) based on individual characteristics, has attracted increasing attention recently in many fields, such as education, economics, and medicine. Current ODR methods usually require the primary outcome of interest in samples for assessing treatment effects, namely the experimental sample. However, in many studies, treatments may have a long-term effect, and as such the primary outcome of interest cannot be observed in the experimental sample due to the limited duration of experiments, which makes the estimation of ODR impossible. This paper is inspired to address this challenge by making use of an auxiliary sample to facilitate the estimation of ODR in the experimental sample. We propose an auGmented inverse propensity weighted Experimental and Auxiliary sample-based decision Rule (GEAR) by maximizing the augmented inverse propensity weighted value estimator over a class of decision rules using the experimental sample, with the primary outcome being imputed based on the auxiliary sample. The asymptotic properties of the proposed GEAR estimators and their associated value estimators are established. Simulation studies are conducted to demonstrate its empirical validity with a real AIDS application.

\end{abstract}

\section{Introduction}\label{sec:1}

Personalized optimal decision making, finding the optimal decision rule (ODR) based on individual characteristics to maximize the mean outcome of interest, has attracted increasing attention recently in many fields. 
Examples include offering customized incentives to increase sales and level of engagement in the area of economics \citep{turvey2017optimal}, developing an individualized treatment rule for patients to optimize expected clinical outcomes of interest in precision medicine \citep{chakraborty2013statistical}, and designing a personalized advertisement recommendation system to raise the click rates in the area of marketing \citep{cho2002personalized}. 

The general setup for finding the ODR contains three components in an experimental sample (from either randomized trials or observational studies): the covariate information ($X$), the treatment information ($A$), and the outcome of interest ($Y$). 
However, current ODR methods cannot be applied to cases where treatments have a long-term effect and the primary outcome of interest cannot be observed in the experimental sample. Take the AIDS Clinical Trials Group Protocol 175 (ACTG 175) data \citep{hammer1996trial} as an example.
 The experiment randomly assigned HIV-infected patients to competitive antiretroviral regimens, and recorded their CD4 count (cells/mm3) and CD8 count over time. A higher CD4 count usually indicates a stronger immune system.
 However, due to the limitation of the follow-up, the clinical meaningful long-term outcome of interest for the AIDS recovery may be missing for a proportion of patients.
Similar problems are also considered in the evaluation of education programs, such as the Student/Teacher Achievement Ratio (STAR) project \citep{word1990state,chetty2011does} that studied long-term impacts of early childhood education on the future income. 
Due to the heterogeneity in individual characteristics, one cannot find a unified best treatment for all subjects. However, the effects of treatment on the long-term outcome of interest can not be evaluated using the experimental data solely. Hence, deriving an ODR to maximize the expected long-term outcome based on baseline covariates obtained at an early stage is challenging.


This paper is inspired to address the challenge of developing ODR when the long-term outcome cannot be observed in the experimental sample. Although the long-term outcome may not be observed in the experimental sample, we could instead obtain some intermediate outcomes (also known as surrogacies or proximal outcomes, $M$) that are highly related to the long-term outcome after the treatment was given. For instance, the CD4 and CD8 counts recorded after a treatment is assigned, have a strong correlation with the healthy of the immune system, and thus can be viewed as intermediate outcomes. 
A natural question is whether an ODR to maximize the expected long-term outcome can be estimated based on the experimental sample (that consists of $\{X,A,M\}$) only. The answer is generally no mainly for two reasons. First, it is common and usually necessary to have multiple intermediate outcomes to characterize the effects of treatment on the long-term outcome. However, when there are multiple intermediate outcomes, it is hard to determine which intermediate outcome or what combination of intermediate outcomes will lead to the best ODR for the long-term outcome. Second, to derive the ODR that maximizes the expected long-term outcome of interest based on the experimental sample, we need to know the relationship between the long-term outcome, intermediate outcomes and baseline covariates, which is generally not practical.  

In this work, we propose using an auxiliary data source, namely the auxiliary sample, to recover the missing long-term outcome of interest in the experimental sample, based on the rich information of baseline covariates and intermediate outcomes. Auxiliary data, such as electronic medical records or administrative records, are now widely accessible. These data usually contain rich information for covariates, intermediate outcomes, and the long-term outcome of interest. However, since they are generally not collected for studying treatment effects, treatment information may not be available in auxiliary data. In particular, in this work, we consider the situation that an auxiliary data consisting of $\{X,M,Y\}$ is available, where Y is the long-term outcome of interest. Note it is also impossible to derive ODR based on such auxiliary sample due to missing treatments.

Our work contributes to the following folds. First, to the best of our knowledge, this is the first work on estimating the heterogeneous treatment effect and developing the optimal decision making for the long-term outcome that cannot be observed in an experiment, by leveraging the idea from semi-supervised learning.  Methodologically, we propose an auGmented inverse propensity weighted Experimental and Auxiliary sample-based decision Rule, named GEAR. This rule maximizes the augmented inverse propensity weighted (AIPW) estimator of the value function over a class of interested decision rules using the experimental sample, with the primary outcome being imputed based on the auxiliary sample. Theoretically, we show that the AIPW estimator under the proposed GEAR is consistent and derive its corresponding asymptotic distribution under certain conditions. A confidence interval (CI) for the estimated value is provided.



There is a huge literature on learning the ODR, including Q-learning \citep{watkins1992q,zhao2009reinforcement,qian2011performance}, A-learning \citep{robins2000,murphy2003optimal,shi2018high}, value search methods \citep{zhang2012robust,zhang2012dynamic,wang2018quantile,nie2020learning}, outcome weighted learning \citep{zhao2012,zhao2015,zhou2017residual}, 
targeted minimum loss-based estimator \citep{van2015targeted}, 
and decision list-based methods \citep{zhangyc2015,zhangyc2016}. 
While none of these methods could derive ODR from the experimental sample with unobserved long-term outcome of interest. 

Our considered estimation of the ODR naturally falls in the framework of semi-supervised learning. A large number of semi-supervised learning methods have been proposed for the regression or classification problems \citep{zhu2005semi, chen2008semiparametric,chapelle2009semi,chakrabortty2018efficient}. Recently, \citet{athey2019surrogate} studied estimation of the average treatment effect under the framework of combining the experimental data with the auxiliary data, where the missing outcomes in the experimental data are imputed based on the regression model learned from the auxiliary data using baseline covariates and intermediate outcomes. However, as far as we know, no work has been done for estimating the ODR in such a semi-supervised setting.

The rest of this paper is organized as follows. We introduce the statistical framework for estimating the optimal treatment decision rule using the experimental sample and the auxiliary sample, and associated assumptions in Section \ref{sec:2}. In Section \ref{sec:3}, we propose our GEAR method and establish consistency and asymptotic distributions of the estimated value functions under the proposed GEAR. Extensive simulations are conducted to demonstrate the empirical validity of the proposed method in Section \ref{sec:4}, followed by an application to ACTG 175 data in Section \ref{:sec5}. We conclude our paper with a discussion in Section \ref{sec:6}. The technical proofs and sensitivity studies under model assumption violation are given in the appendix.

\section{Statistical Framework} \label{sec:2}
\subsection{Experimental Sample and Auxiliary Sample}
Suppose there is an experimental sample of interest $E$. Let $X_E$ denote $r$-dimensional individual's baseline covariates with the support $\mathbb{X}_E \in \mathbb{R}^r$, and $A_E\in\{0,1\}$ denote the treatment an individual receives. The long-term outcome of interest $Y_E $ with support $\mathbb{Y}_E \in \mathbb{R}$ cannot be observed, instead we only obtain the $s$-dimensional intermediate outcomes $M_E$ with support $\mathbb{M}_E \in \mathbb{R}^s$ after a treatment $A_E$ is assigned. Denote $N_E$ as the sample size for the experimental sample, which consists of $\{E_i=(X_{E,i},A_{E,i},M_ {E,i}), i = 1, \dots , N_E\}$ independent and identically distributed (I.I.D.) across $i$. 

To recover the missing long-term outcome of interest in the experimental sample, we include an auxiliary sample, $U$, which contains the individual's baseline covariates $X_U$, intermediate outcomes $M_U$, and the observed long-term outcome of interest  $Y_U$, with support $\mathbb{X}_U,\mathbb{M}_U,\mathbb{Y}_U$ respectively. However, treatment information is not available in the auxiliary sample. Let $N_U$ denote the sample size for the I.I.D. auxiliary sample that includes $\{U_i=(X_{U,i},M_{U,i},Y_ {U,i}), i = 1, \dots , N_U\}$. 

We use $R=\{E,U\}$ to indicate the missingness and identification of each sample, where $R=E$ implies the experimental sample with missing long-term primary outcome and $R=U$ means the auxiliary sample with missing treatment information. Thus, these two samples can also be rewritten as one joint sample $\{(X_{i},R_{i},A_i \mathbb{I}_{R_{i}=E},M_{i},Y_ {i}\mathbb{I}_{R_{i}=U}), i = 1, \dots , N_E+N_U\}$, where $\mathbb{I}(\cdot)$ is an indicator function.


\subsection{Assumptions}
In this subsection, we make five key assumptions in order to introduce the ODR. For the experimental sample, define the potential outcomes $Y_E^*(0)$ and $Y_E^*(1)$ as the long-term outcome that would be observed after an individual receiving treatment 0 or 1, respectively. Let the propensity score as the conditional probability of receiving treatment 1 in the experimental sample, i.e. $\pi(x)=Pr_E(A_{E,i}=1|X_{E,i}=x)$. As standard in causal inference by \citet{rubin1978bayesian}, we assume:

\noindent (A1). Stable Unit Treatment Value Assumption (SUTVA): 
$
Y_E= A_E Y_E^{\star}(1)  + (1-A_E)Y_E^{\star}(0).
$

\noindent (A2). No Unmeasured Confounders Assumption: 
$
\{Y_E^*(0),Y_E^*(1)\}\independent A_E\mid X_E.
$

\noindent (A3). $0<\pi(x)<1$ for all $x \in \mathbb{X}_E$.

To impute the missing long-term outcome in the experimental sample with the assistance of the auxiliary sample, we introduce the following two assumptions, the comparability assumption and the surrogacy assumption. 

First, the comparability assumption states that the population distribution of the long-term outcome of interest $Y$ is independent of whether belonging to the experimental sample or the auxiliary sample, given the information of population baseline covariates $X$ and population intermediate outcomes $M$ as follows.

\noindent (A4). Comparability Assumption: 
$Y \independent R\mid X,M$.

Here, (A4) is also known as `conditional independence assumption' made in \citet{chen2008semiparametric}, and has an equivalent expression as $Y_E \mid \{M_E,X_E\} \sim Y_U \mid \{M_U,X_U\}$ proposed in \citet{athey2019surrogate}. When (A4) holds, we have a direct conclusion of the equality of the conditional mean outcome given baseline covariates and intermediate outcomes in each sample, stated in the following corollary.

\begin{coro}{(Equal Conditional Mean)}\label{coro1}
Under (A4),
\begin{eqnarray}\label{equalmean}
\mathbb{E}[Y_E|M_E=m,X_E=x]=\mathbb{E}[Y_U|M_U=m,X_U=x].
\end{eqnarray}
\end{coro}
\begin{remark}
It is shown in Section \ref{sec:3} that (A4) can be relaxed to Equation (\ref{equalmean}) for deriving the proposed method. 
\end{remark}

We further define the missing at random (MAR) assumption in the joint sample as: 
$
\{Y,A\} \independent R\mid X,M;
$
and give the following corollary to show the relationship between (A4) and the MAR assumption.

\begin{coro}{(MAR Assumption)}\label{coro2}
\begin{eqnarray*}
\{Y,A\} \independent R\mid X,M \longrightarrow Y \independent R\mid X,M.
\end{eqnarray*}
\end{coro}

\begin{remark}
Corollary \ref{coro2} is a direct result of joint independence implying marginal independence. Though (A4) is untestable due to the missing long-term outcome in the experimental sample, one can believe (A4) holds if there exists strong evidence about the reasonability of the MAR assumption in the joint sample.
\end{remark}


Second, the surrogacy assumption states that the long-term outcome of interest in the experimental sample is independent of the treatment conditional on a set of baseline covariates and intermediate outcomes as below.


\noindent (A5). Surrogacy Assumption: 
$
Y_E \independent A_E\mid X_E,M_E.
$
\begin{remark}
The above assumption is also used in \citet{athey2019surrogate}. 
The validation of the surrogacy assumption relies on the `richness' of intermediate outcomes that are highly related to the long-term outcome of interest. Similarly, it is infeasible to check the surrogacy assumption due to the missing long-term outcome in the experimental sample.
\end{remark}

\begin{figure} 
\centering
\includegraphics[width=3in]{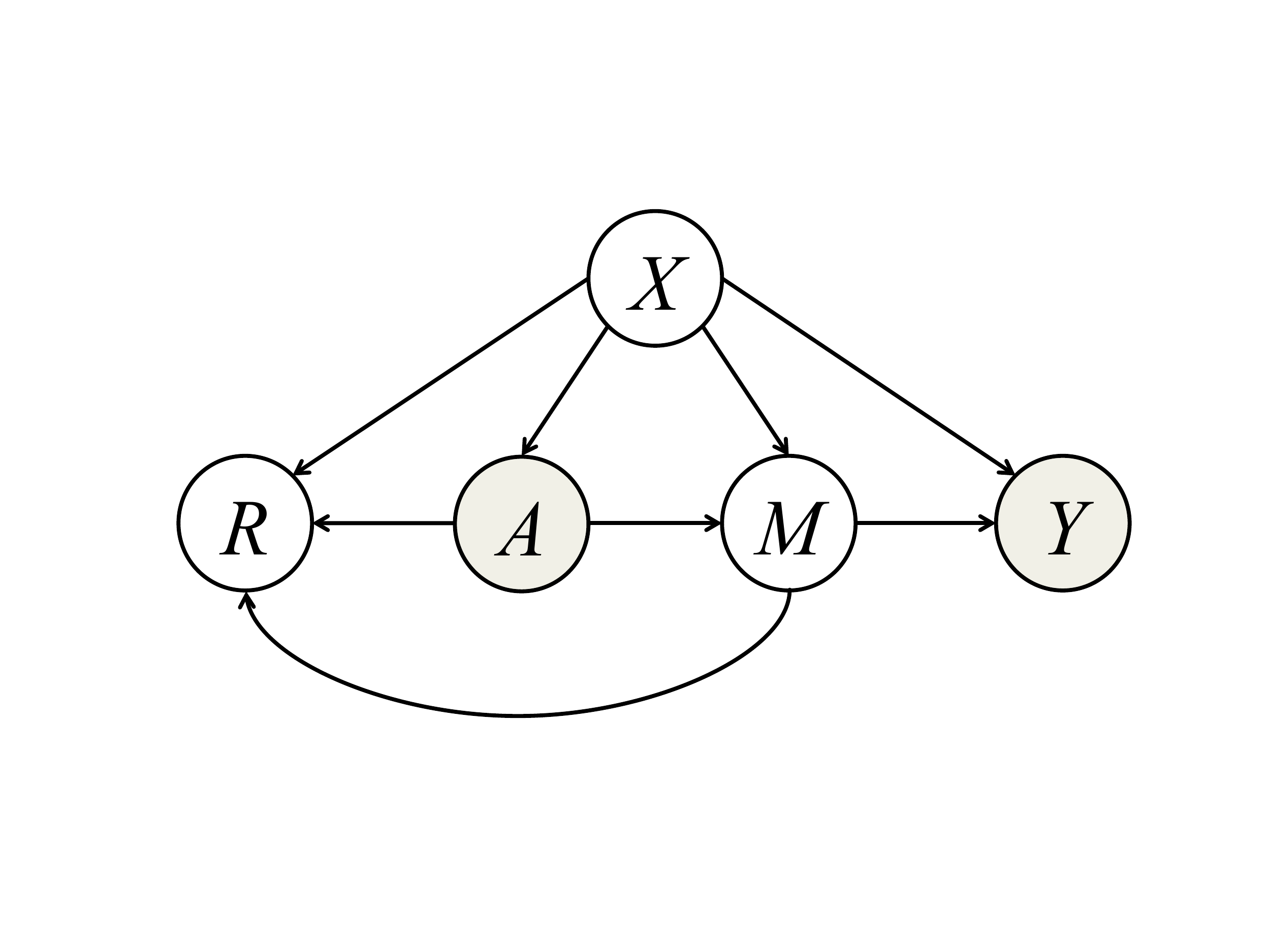}

\caption{A direct acyclic graph illustrating assumptions (A2), (A4), and (A5) in the joint sample. White nodes represent observed variables, and grey nodes are variables with missing values.}\label{dag}
\end{figure}

We illustrate the statistical framework of the joint sample under above assumptions by a direct acyclic graph in Figure \ref{dag}. Graphically, $A$ and $Y$ have no common parents except for $X$, encoding (A2); $R$ and $Y$ have two common parents, $X$ and $M$, encoding (A4); when fixing $X$ and $M$, $A$ and $Y$ are independent, encoding (A5).


\subsection{Value Function and Optimal Decision Rule}
A decision rule is a deterministic function $d(\cdot)$ that maps $\mathbb{X}_E$ to $\{0,1\}$. Define the potential outcome of interest under $d(\cdot)$ as 
$
	Y_E^*(d)=Y_E^*(0) \{1-d(X_E)\}+Y_E^*(1) d(X_E),
	$
which would be observed if a randomly chosen individual from the experimental sample had received a treatment according to $d(\cdot)$, where we suppress the dependence of $Y_E^*(d)$ on $X_E$. We then define the value function under $d(\cdot)$ as the expectation of the potential outcome of interest over the experimental sample as 
\begin{eqnarray*} 
 V(d)=\Mean \{Y_E^*(d)\}=\Mean [Y_E^*(0) \{1-d(X_E)\}+Y_E^*(1) d(X_E)]. 
\end{eqnarray*} 
As a result, we have the optimal treatment decision rule (ODR) of interest defined to maximize the value function over the experimental sample among a class of decision rules of interest as
$
d^{opt}(\cdot)=\argmin_{d(\cdot)} V(d).
$
Suppose the decision rule $d(\cdot)$ relies on a model parameter $\beta$, denoted as $d(\cdot)\equiv d(\cdot;\beta)$. We use a shorthand to write $V(d)$ as $V(\beta)$, and define
$
\beta_0 =\argmin_{\beta} V(\beta).
$
Thus, the value function under the true ODR $d(\cdot;\beta_0)$ is defined as $V(\beta_0)$.

\section{Proposed Method}\label{sec:3}
In this section, we detail the proposed method by constructing the AIPW value estimator for the long-term outcome based on two samples. Implementation details are provided to find the ODR. The consistency and asymptotical distribution of the value estimator under our proposed GEAR are presented, followed by its confidence interval. All the proofs are provided in Section B of the appendix. 

\subsection{AIPW Estimator for Long-Term Outcome}


To overcome the difficulty of estimating the value function due to the missing long-term outcome of interest in the experimental sample, one intuitive way is to impute the missing outcome $Y_E$ with its conditional mean outcome given baseline covariates and intermediate outcomes (total common information available in both samples). 

Denote $\mu_E(m,x)\equiv \Mean[Y_E|M_E=m,X_E=x]$, and $\mu_U(m,x)\equiv \Mean[Y_U|M_U=m,X_U=x]$. Under Corollary \ref{coro1}, we have $\mu_E(m,x)=\mu_U(m,x)$. Here, $\mu_E(m,x)$ is inestimable because of the missing long-term outcome. We instead use $\mu_U(M_E,X_E)$ to impute the missing $Y_E$ and give the following lemma as a middle step to construct the AIPW value estimator for the long-term outcome. 

\begin{lemma}\label{lem1}
Under (A1)-(A5), given $d(\cdot;\beta)$, we have
\begin{eqnarray*}
V(\beta)=\Mean \Bigg[{\mathbb{I}\{A_E=d(X_E;\beta)\}\mu_U(M_E,X_E)\over{A_E\pi(X_E)+(1-A_E)\{1-\pi(X_E)\}}}\Bigg].
\end{eqnarray*}
\end{lemma}

Next, we propose the AIPW estimator of the value function for the long-term outcome in the experimental sample. To address the difficulty of forming the augmented term when the long-term outcome of interest cannot be observed, we show that augmenting on the missing long-term outcome is equivalent to augmenting on the imputed conditional mean outcome of interest $\mu_U(M_E,X_E)$, by the following lemma. 

\begin{lemma}\label{lem2}
Under (A1)-(A5), given $d(\cdot;\beta)$, we have
\begin{eqnarray*}
 \Mean_{Y_E|X_E}\{Y_E|A_E=d(X_E;\beta),X_E\} 
= \Mean_{M_E|X_E}\{\mu_U(M_E,X_E)|A_E=d(X_E;\beta),X_E\},
\end{eqnarray*}
where $E_{A|B}$ means taking expectation with respect to the conditional distribution of $A$ given $B$.
\end{lemma}

According to Lemma \ref{lem1} and Lemma \ref{lem2}, given a decision rule $d(\cdot;\beta)$, the value function $V(\beta)$ can be consistently estimated through
\begin{eqnarray*}
\begin{split}
V^{\star}_{n,AIP}(\beta)=  & {1\over N_E}\sum_{i=1}^{N_E} \Bigg[\nu_i +
{\mathbb{I}\{A_{E,i}=d(X_{E,i};\beta)\}\{{\mu}_U(M_{E,i},X_{E,i})-\nu_i\}\over{A_{E,i} \pi (X_{E,i})+(1-A_{E,i})\{1-\pi(X_{E,i})\}}}\Bigg], 
\end{split} 
\end{eqnarray*}
where $\nu_i \equiv \Mean\{\mu_U(M_{E,i},X_{E,i})|A_{E,i}=d(X_{E,i};\beta),X_{E,i}\}$ presents the augmented term. Here, the propensity score $\pi$ can be estimated in the experimental sample, denoted as $\widehat{\pi}$, and the conditional mean $\mu_U$ can be estimated in the auxiliary sample, denoted as $\widehat{\mu}_U$. 
Then, by replacing the implicit functions in $V^{\star}_{n,AIP}(\beta)$, it is straightforward to give the AIPW estimator of the value function $V(\beta)$ as
\begin{eqnarray*} 
\begin{split}
\widehat{V}_{AIP}(\beta) = & {1\over N_E}\sum_{i=1}^{N_E} \Bigg[\widehat{\nu}_i+{\mathbb{I}\{A_{E,i}=d(X_{E,i};\beta)\}\{\widehat{\mu}_U(M_{E,i},X_{E,i})- \widehat{\nu}_i\}\over{A_{E,i}\widehat{\pi}(X_{E,i})+(1-A_{E,i})\{1-\widehat{\pi}(X_{E,i})\}}}\Bigg] ,
\end{split} 
\end{eqnarray*}
where  $\widehat{\nu}_i \equiv \widehat{\Mean}\{\widehat{\mu}_U(M_{E,i},X_{E,i})|A_{E,i}=d(X_{E,i};\beta),X_{E,i}\}$ is the estimator for $\nu_i$.
We define $\widehat{\beta}^{G}={\argmax}_\beta \widehat{V}_{AIP}(\beta)$, and then propose the GEAR as $d(X;\widehat{\beta}^{G})$ with the corresponding estimated value function as $\widehat{V}_{AIP}(\widehat{\beta}^{G})$. 
  


\subsection{Implementation Details}\label{sec:impl}
\textbf{Class of decision rules:}  
The GEAR can be searched within a pre-specified class of decision rules. Popular classes include generalized linear rules, fixed depth decision trees, threshold rules, and so on \citep{zhang2012robust,athey2017efficient,rai2018statistical}. In this paper, we focus on the class of generalized linear rules. Specifically, suppose the decision rule takes a form as $d(X_E;\beta) \equiv \mathbb{I}\{g(X_E)^\top  \beta > 0\}$, where $g(\cdot)$ is an unknown function. We use $\phi_X(\cdot)$ to denote a set of basis functions of $\mathbb{X}_E$ with length $v$, which are ``rich'' enough to approximate the underlying function $g(\cdot)$. Thus, the GEAR is found within a class of $\mathbb{I}\{\phi_X(X_E)^\top  \beta > 0\}$. For notational simplicity, we include 1 in $\phi_X(\cdot)$ so that $\beta \in  \mathbb{R}^{v+1}$. With subject to $||\beta||_2=1$ for identifiability purpose, the maximizer for $ \widehat{V}_{AIP}(\beta)$ can be solved using any global optimization algorithm. In our implementation, we apply the heuristic algorithm to search for the GEAR.


 \textbf{Estimation models:} 
The conditional mean of the long-term outcome $\mu_U(m,x)$ can be estimated through any parametric or nonparametric model. In practice, we assume $\mu_U(m,x)$ can be determined by a flexible basis function of baseline covariates and intermediate outcomes, to fully capture the underlying true model. Similarly, one can use a flexible basis function of baseline covariates and the treatment to model the augmented term as well as the propensity score function. Note that any machine learning tools such as Random Forest or Deep Learning can be applied to model terms in the proposed AIPW estimator. Our theoretical results still hold under these nonparametric models as long as the regressors have desired convergence rates (see results established in \citet{wager2018estimation,farrell2018deep}).

\textbf{Estimation of the augmented term:} 
To estimate the augmented term $\nu_i$,
we need three steps as follows. First, we model $\mu_U(m,x)$ through the auxiliary sample $\{X_U,M_U,Y_U\}$ as $\widehat{\mu}_U(m,x)$; second, we plug $\{M_E,X_E\}$ of the experimental sample into $\widehat{\mu}_U(m,x)$ and get $\widehat{\mu}_U(M_{E},X_{E})$ as the conditional mean outcome of interest to impute the missing $Y_E$; at last, we fit $\widehat{\mu}_U(M_{E},X_{E})$ on $\{A_E,X_E\}$ in the experimental sample, and get $\widehat{\nu}_i$.

\subsection{Theoretical Properties}\label{sec:thms}

We next show the consistency and asymptotic normality of our proposed AIPW estimator. Its asymptotic variance can be decomposed into two parts, corresponding to the estimation variances from two independent samples. 
As mentioned in Section \ref{sec:impl}, our AIPW estimator can handle various machine learning or parametric estimators as long as regressors have desired convergence rates. To derive an explicit variance form, we next focus on parametric models. 

We posit parametric models for $\pi (x)\equiv \pi (x;\gamma)$ and $\mu_U(m,x)\equiv \mu_U(m,x;\lambda)$ with true model parameters $\gamma$ and $\lambda$. 
Let $\phi_X(X)$ and $\phi_M(M)$ to represent appropriate basis functions for $X$ and $M$, respectively. Without loss of generality, we posit basis model for the augmented term such that $\Mean\{\mu_U(m,x;\lambda)|A=0,X=x\}\equiv \phi_X(x)^\top \theta_0$, and $\Mean\{\mu_U(m,x;\lambda)|A=1,X=x\}\equiv  \phi_X(x)^\top \theta_1$ with true model parameters $\theta_0$ and $\theta_1$. The following conditions are needed to derive our theoretical results:\\
(A6). Suppose the density of covariates $f_X(x)$ is bounded away from 0 and $\infty$ and is twice continuously differentiable with bounded derivatives.\\
(A7). Both $ \pi (x;\gamma)$ and $ \mu_U(m,x;\lambda)$ are smooth bounded functions, with their first derivatives exist and bounded.\\
(A8). Model for $ \mu_U(m,x;\lambda)$ is correctly specified. \\
%
(A9). Denote $t=\sqrt{{N_E\over N_U}}$
and assume $0<t<+\infty$. \\
(A10). The true value function $V(\beta)$ is twice continuously differentiable at a neighborhood of $\beta_0$.  \\
(A11). Either the model of the propensity score or the model of the augmented term is correctly specified.

 Here, (A6) and (A10) are commonly imposed to establish the inference for value search methods \citep{zhang2012robust,wang2018quantile}. (A7) is assumed for desired convergence rates of $\widehat{\pi}$ and $\widehat{\mu}_U$. To apply machine learning tools, similar assumption is required (see more details in \citet{wager2018estimation,farrell2018deep}). From (A8), we can replace the missing long-term outcome with its imputation, and thus the consistency holds. Evaluations are provided in Section \ref{misspe} to examine the proposed method when (A8) is violated. (A9) states that the sizes of two samples are comparable, which prevents the asymptotic variance from blowing up when combining two samples in semi-supervised learning \citep{chen2008semiparametric,chakrabortty2018efficient}. (A11) is included to establish the doubly robustness of the value estimator, which is commonly used in the literature of doubly robust estimator \citep{dudik2011doubly,zhang2012robust,zhang2012dynamic}.
  


The following theorem gives the consistency of our AIPW estimator of the value function to the true value function. 
\begin{thm}{(Consistency)}\label{thm3}
Under (A1)-(A9) and (A11), 
\begin{eqnarray*}
\widehat{V}_{AIP}(\beta) =V(\beta)+o_p(1), \quad \forall \beta.
\end{eqnarray*}
\end{thm}

\begin{remark}
When the model for $\mu_U(m,x)$ is correctly specified, our AIPW estimator is doubly robust given either the model of the propensity score or the model of the augmented term is correct. To prove the theorem, we establish the theoretical results with their proofs for the inverse propensity-score weighted estimator as a middle step. See more details in Section A of the appendix.
\end{remark}

To establish the asymptotic normality of $\widehat{V}_{AIP}(\widehat{\beta}^{G})$, we first show the estimator $\widehat{\beta}^{G}$ has a cubic rate towards the true $\beta_0$. 
\begin{lemma}\label{lem3}
Under (A1)-(A11), we have
\begin{eqnarray}\label{cubic}
N_E^{1/3}||\widehat{\beta}^{G}-\beta_0||_2 = O_p(1),
\end{eqnarray}
where $||\cdot||_2$ is the $L_2$ norm, and $O_p(1)$ means the random variable is stochastically bounded.
\end{lemma}

Based on Lemma \ref{lem3}, we next give the asymptotic normality of $\sqrt{N_E}\big\{\widehat{V}_{AIP}(\widehat{\beta}^{G})-V(\beta_0)\big\}$ in the following theorem.
\begin{thm}{(Asymptotic Distribution)}\label{thm4}
Under (A1)-(A11), 
\begin{eqnarray}\label{asyaipw}
\sqrt{N_E}\big\{\widehat{V}_{AIP}(\widehat{\beta}^{G})-V(\beta_0)\big\} \overset{\mathcal{D}}{\longrightarrow} N(0,\sigma_{AIP}^2),
\end{eqnarray}
where $\sigma_{AIP}^2=t\sigma_{U}^2+\sigma_{E}^2$, $\sigma_{U}^2=\Mean[\{\xi_{i}^{({U})}\}^2]$, and $\sigma_{E}^2=\Mean[\{\xi_{i}^{(E)}\}^2]$. Here, $\xi_{i}^{(E)}$ and $\xi_{i}^{(U)}$ are the I.I.D. terms in the experimental sample and auxiliary sample, respectively.
\end{thm}

\begin{remark}
From Theorem \ref{thm4}, the asymptotic variance of the AIPW estimator has an additive form that consists of the estimation error from each sample. Proportion of these two estimation variances is controlled by the sample ratio. In reality, $N_U$ is usually larger than $N_E$. When $N_U/N_E\to \infty$, we have $t\to 0$, and thus the estimation error from auxiliary sample can be ignored. Our result under this special case is supported by \citet{chakrabortty2018efficient} where they considered $N_U/N_E\to \infty$ for a regression problem. 
\end{remark}
 
Next, we give explicit form of $\xi_{i}^{(E)}$ and $\xi_{i}^{(U)}$ from the proof of Theorem \ref{thm4} to estimate $\sigma_{AIP}$. Denote $\dot{\pi} (x;\gamma) \equiv \partial \pi (x;\gamma)/\partial\gamma$ and $\dot{\mu}_U(m,x;\lambda)\equiv \partial\mu_U(m,x;\lambda)/\partial \lambda.$  Let 
\begin{eqnarray*}
H_1&&\equiv \underset{N_E\to +\infty}{\lim} {1\over N_E}\sum_{i=1}^{N_E} \phi_X(X_{E,i}) \dot{\pi} (X_{E,i};\gamma)^\top  , \\
H_2&&\equiv \underset{N_U\to +\infty}{\lim} {1\over N_U}\sum_{i=1}^{N_U} \begin{bmatrix}
   			\phi_X(X_{U,i})\\
   			\phi_M(M_{U,i})
		\end{bmatrix}\dot{\mu}_U(M_{U,i},X_{U,i};\lambda)^\top  , \\
H_3&&\equiv \underset{N_E\to +\infty}{\lim} {1\over N_E}\sum_{i=1}^{N_E} (1-A_{E,i})\phi_X(X_{E,i})\phi_X( X_{E,i})^\top  ,\\
 H_4&&\equiv \underset{N_E\to +\infty}{\lim} {1\over N_E}\sum_{i=1}^{N_E} A_{E,i}\phi_X(X_{E,i}) \phi_X(X_{E,i})^\top  , \\
 G_1&&\equiv \underset{N_E\to +\infty}{\lim} {1\over N_E}\sum_{i=1}^{N_E} {r_i (1-2A_{E,i})\dot{\pi}(X_{E,i};\gamma) {\mu}_U(M_{E,i},X_{E,i};\lambda)\over{s_i^2}}, \\
 G_2&& \equiv \underset{N_E\to +\infty}{\lim} {1\over N_E}\sum_{i=1}^{N_E} {r_i\over{s_i}} \dot{\mu}_U(M_{E,i},X_{E,i};\lambda) , \\
 G_3&&\equiv \underset{N_E\to +\infty}{\lim} {1\over N_E}\sum_{i=1}^{N_E} - {r_i q_i (1-2A_{E,i})\dot{\pi} (X_{E,i};\gamma) \over{s_i^2}}, \\
 G_4&& \equiv \underset{N_E\to +\infty}{\lim} {1\over N_E}\sum_{i=1}^{N_E}\Big[1- {r_i\over s_i} \Big]\phi_X(X_{E,i})\{1-d(X_{E,i};\beta_0)\},  \\
 G_5&&\equiv \underset{N_E\to +\infty}{\lim} {1\over N_E}\sum_{i=1}^{N_E}\Big[1- {r_i\over s_i} \Big]\phi_X(X_{E,i})d(X_{E,i};\beta_0),  
 \end{eqnarray*}
where $r_i \equiv \mathbb{I}\{A_{E,i}=d(X_{E,i};\beta_0)\}$, $s_i \equiv A_{E,i} \pi(X_{E,i};\gamma) +(1-A_{E,i})\{1-\pi(X_{E,i};\gamma)\}$, and $q_i \equiv \phi_X(X_{E,i})^\top \theta_0+\phi_X(X_{E,i})^\top (\theta_1-\theta_0)d(X_{E,i};\beta_0)$. 
%
%

Then, the I.I.D. term in the experimental sample is
\begin{eqnarray*} \begin{split}
 \xi_{i}^{(E)}\equiv &{r_i \{{\mu}_U(M_{E,i},X_{E,i};\lambda)-\nu_{i}^*\}\over{s_i}}+\nu_{i}^*-V(\beta_0)\\
&+(G_1^\top +G_3^\top ) H_1^{-1} \phi_X(X_{E,i})\{A_{E,i}-\pi (X_{E,i};\gamma)\}\\
&+G_5^\top  H_4^{-1}  \phi_X(X_{E,i})A_{E,i}\{{\mu}_U(M_{E,i},X_{E,i};\lambda)- \phi_X(X_{E,i})^\top  \theta_1\}\\
&+G_4^\top  H_3^{-1} \phi_X(X_{E,i})(1-A_{E,i}) \{{\mu}_U(M_{E,i},X_{E,i};\lambda)- \phi_X(X_{E,i})^\top  \theta_0\},
\end{split} 
\end{eqnarray*}
for $\nu_{i}^* \equiv E\{\mu_U(M_{E,i},X_{E,i};\lambda)|A_{E,i}=d(X_{E,i};\beta_0),X_{E,i}\}$. 
And \begin{eqnarray*} \xi_{i}^{(U)}\equiv G_2^\top H_2^{-1}\begin{bmatrix}
   			\phi_X(X_{U,i})\\
   			\phi_M(M_{U,i})
		\end{bmatrix} \{Y_{U,i}-{\mu}_U(M_{U,i},X_{U,i};\lambda)\}, \end{eqnarray*}
		 corresponds to the I.I.D. term in the auxiliary sample. 


By plugging the estimations 
into the pre-specified models, 
we could obtain the estimated $\widehat{\xi_{i}}^{(E)}$ and $\widehat{\xi_{i}}^{(U)}$. Then the variance $\sigma_{E}^2$ and $\sigma_{U}^2$ can be consistently estimated by $\widehat{\sigma}_{E}^2={1\over N_E}\sum_{i=1}^{N_E} \{\widehat{\xi_{i}}^{(E)}\}^2$ and $\widehat{\sigma}_{U}^2={1\over N_U}\sum_{i=1}^{N_U} \{\widehat{\xi_{i}}^{(U)}\}^2$, respectively. Thus, we can estimate $\sigma_{AIP}$ through
\begin{eqnarray}\label{sdaipw}
\widehat{\sigma_{AIP}}\equiv \sqrt{t\widehat{\sigma}_{U}^2+\widehat{\sigma}_{E}^2},
\end{eqnarray}
based on Theorem \ref{thm4}. Therefore, a two-sided $1-\alpha$ confidence interval (CI) for $V(\beta_0)$ under the GEAR is 
\begin{eqnarray}\label{ciaipw}
\Big [ \widehat{V}_{AIP}(\widehat{\beta}^{G})-{z_{\alpha/2}\widehat{\sigma_{AIP}}\over\sqrt{N_E}},~~~ \widehat{V}_{AIP}(\widehat{\beta}^{G})+{z_{\alpha/2}\widehat{\sigma_{AIP}}\over\sqrt{N_E} }\Big],
\end{eqnarray}
where $z_{\alpha/2}$ denotes the upper $\alpha/2-$th quantile of a standard normal distribution.

\section{Simulation Studies}\label{sec:4}
In this section, we evaluate the proposed method 
when the model of the conditional mean of the long-term outcome correctly specified and misspecified separately in the following two subsections. Additional sensitivity studies when assumptions are violated are provided in Section C of the appendix. 

\subsection{Evaluation under Correctly Specified Model}\label{value_infer}
Simulated data, including baseline covariates $X=[X^{(1)},X^{(2)},\cdots,X^{(r)}]^\top $, the treatment $A$, intermediate outcomes $M=[M^{(1)},M^{(2)},\cdots,M^{(s)}]^\top$, and the long-term outcome $Y$, are generated from the following model:
\begin{eqnarray*}  
\begin{split}
&X^{(1)},X^{(2)},\cdots,X^{(r)} \overset{iid}{\sim} Uniform[-1,1], \quad A \overset{iid}{\sim} Bernoulli (0.5),\\
&M=H^M(X)+ A C^M(X)+\epsilon^M,  ~~~Y=H^Y(X)+C^Y(X, M)+\epsilon^Y,
\end{split} 
\end{eqnarray*}
where $\epsilon^M$ and $\epsilon^Y$ are random errors following $N(0,0.5)$. Here, $A$ in the auxiliary sample is used only for generating intermediate outcomes such that the comparability assumption is satisfied. Note that $Y$ is generated for the auxiliary sample only. Given $X$ and $M$,
we can see $Y$ is independent of $A$, which indicates the surrogacy assumption. 

Set $r = 4$ and $s=2$. We consider following two scenarios with different $H^M(\cdot)$, $C^M(\cdot)$, $H^Y(\cdot)$, and $C^Y(\cdot)$. 
 \begin{eqnarray*} 
 \textbf{S1}:
	\left\{\begin{array}{ll}
		H^M(X)=
		\begin{bmatrix}
   			X^{(3)}\\
   			X^{(1)}
		\end{bmatrix},
		C^M(X)=
		\begin{bmatrix}
   			4\{X^{(1)}-X^{(2)}\}\\
   			4\{X^{(4)}-X^{(3)}\}
		\end{bmatrix},\\
		H^Y(X)=-1+X^{(2)}+X^{(4)},
		C^Y(X, M)=M^{(1)}+M^{(2)}.\\
	\end{array} 
	\right.\\
	\end{eqnarray*}
	 \begin{eqnarray*} 
\textbf{S2}:
	\left\{\begin{array}{ll}
		H^M(X)=
		\begin{bmatrix}
   			\{X^{(1)}\}^2X^{(3)}+\sin\{X^{(4)}\}\\
   			\{X^{(1)}\}^3-\{X^{(2)}-X^{(4)}\}^2
		\end{bmatrix},\\
		C^M(X)=
		\begin{bmatrix}
   			4\{X^{(1)}-X^{(2)}\}\\
   			4\{X^{(4)}-X^{(3)}\}
		\end{bmatrix},\\
		H^Y(X)=-1+X^{(2)}+X^{(4)},
		C^Y(X, M)=M^{(1)}+M^{(2)}.\\
	\end{array}
	\right. 
\end{eqnarray*}
Under Scenario 1 and 2, we have the parameter of the true ODR as $\beta_0=[0,0.5,-0.5,-0.5,0.5]^\top $ with subject to $||\beta_0||_2=1$, which can be easily solved based on the function $C^M(\cdot)$ that describes the treatment-covariates interaction. The true value $V(\beta_0)$ can be calculated by Monte Carlo approximations, as listed in Table \ref{table:2}. 
We consider $N_U=400$ for the auxiliary sample and allow $N_E$ chosen from the set $\{200,400,800\}$ in the experimental sample. 

To apply the GEAR, we model the conditional mean of the long-term outcome $\mu_U(m,x)$ and the augmented term $v_i$ in the auxiliary data via a linear regression. Here, the model of $\mu_U(m,x)$ is correctly specified by noting that $Y$ is linear in $\{X,M\}$ under Scenario 1 and 2. The GEAR is searched within a class of $d(X_E;\beta) =\mathbb{I}(X_E^\top  \beta > 0)$ subjecting to $||\beta||_2=1$, through Genetic Algorithm provided in \proglang{R} package \code{rgenound}, where we set `optim.method' = `Nelder-Mead', `pop.size' = 3000, `domain'=[-10,10], and `starting.values' as a zero vector.
Results are summarized in Table \ref{table:2}, including 
the estimated value under the estimated rule $\widehat{V}_{AIP}(\widehat{\beta}^{G})$ and its standard error $SE\{\widehat{V}_{AIP}\}$, the estimated standard deviation $\Mean\{\widehat{\sigma}_{AIP}\}$ by Equation (\ref{sdaipw}), the value under the estimated rule $V(\widehat{\beta}^G)$ by plugging the GEAR into the true model, the empirical coverage probabilities (CP) for 95\% CI constructed by Equation (\ref{ciaipw}), the rate of the correct decision (RCD) made by the GEAR, and the $L_2$ loss of $\widehat{\beta}^G$ ($||\widehat{\beta}^G -\beta_0||_2$), aggregated over 500 simulations.

\begin{table}[t]
\caption{Empirical results 
under the GEAR for Scenario 1 and 2.}\label{table:2} 
\begin{center}
\begin{small}
\begin{sc}
\begin{tabular}{cccc|ccc}
\toprule
&\multicolumn{3}{c|}{Scenario 1} &\multicolumn{3}{c}{Scenario 2} \\
\midrule 
$N_E=$& $200$ & $400$&$800$& $200$ & $400$&$800$\\
\midrule 
\midrule 
$V(\beta_0)$& & 0.87 &  &&0.20& \\
\midrule 
$\widehat{V}_{AIP}(\widehat{\beta}^{G})$&0.89&0.89 &0.88&0.24 &0.24 &0.22 \\
\midrule
 $SE\{\widehat{V}_{AIP}\}$&0.02 &0.01&0.01& 0.02& 0.01 & 0.01 \\
\midrule
 $\Mean\{\widehat{\sigma}_{AIP}\}$&0.02&0.01&0.01&0.02&0.01&0.01\\
\midrule
$V(\widehat{\beta}^G)$ &0.85&0.86&0.86&0.18& 0.18&0.19\\
\midrule
CP (\%) &94.6 &94.8 &94.8 &95.0 &94.4 &94.8 \\
\midrule
RCD (\%) &95.9 &96.6 &97.3&95.0 &95.8 &96.7 \\
\midrule
 $||\widehat{\beta}^G -\beta_0||_2$  & 0.12& 0.09&0.07&0.14& 0.11&0.09\\
\bottomrule

\end{tabular}
\end{sc}
\end{small}
\end{center} 
\end{table}
%
%
 
From Table \ref{table:2}, it is clear that both the estimated GEAR and its estimated value approach to the true as the sample size $N_E$ increases in all scenarios. Specifically, our proposed GEAR method achieves $V(\widehat{\beta}^G)=0.86$ in Scenario 1 ($V( {\beta}_0)=0.87$) and $V(\widehat{\beta}^G)=0.19$ in Scenario 2 ($V( {\beta}_0)=0.20$) when $N_E=800$. Notice that the $\ell_2$ loss of $\widehat{\beta}^G$ decays at a rate that is approximately proportional to $N_E^{-1/3}$, which verifies our theoretical findings in Lemma \ref{lem3}. Moreover, the average rate of the correct decision made by the GEAR increases with $N_E$ increasing. In addition, there are two findings that help to verify Theorem \ref{thm4}. First, the estimated standard deviation of value function is close to the standard error of the estimated value function, and gets smaller as the sample size $N_E$ increases. Second, the empirical coverage probabilities of the proposed 95\% CI approach to the nominal level in all settings. Note that there is no strictly increasing trend of the empirical coverage probabilities due to the fixed sample size $N_U=400$. 
 

\subsection{Evaluation under Model Misspecification}\label{misspe}
We consider more general settings to examine the proposed method when the model of $\mu_U(m,x)$
is misspecified. The data is generated from the same model in Section \ref{value_infer}.
We fix 
  $
		H^M(X)=
		\begin{bmatrix}
   			X^{(3)}\\
   			X^{(1)}
		\end{bmatrix},
		C^M(X)=
		\begin{bmatrix}
   			4\{X^{(1)}-X^{(2)}\}\\
   			4\{X^{(4)}-X^{(3)}\}
		\end{bmatrix},
$ 
and set following three scenarios with different $H^Y(\cdot)$ and $C^Y(\cdot)$.
\begin{eqnarray*} 
\begin{split}
	&\textbf{S3}:\left\{\begin{array}{ll}
		H^Y(X)=\{X^{(1)}+X^{(3)}\}\{X^{(1)}\}^2\\
		~~~~~~~~~~~~~~~~~~~~~~~ ~+\sin\{X^{(4)}\}-\{X^{(2)}-X^{(4)}\}^2,\\
		C^Y(X, M)=M^{(1)}+M^{(2)}.
	\end{array}
	\right.
	\\
\\
&\textbf{S4}:\left\{\begin{array}{ll}
		H^Y(X)=\{X^{(1)}\}^3+\{X^{(2)}\}^2+X^{(3)},\\
		C^Y(X, M)=M^{(1)}+X^{(4)}M^{(2)}.
	\end{array}
	\right.
	\\
\\
&\textbf{S5}:
	\left\{\begin{array}{ll}
		H^Y(X)=X^{(2)}-\{X^{(4)}\}^2,\\
		C^Y(X, M)=0.25\{M^{(1)}-X^{(3)}\}^2+M^{(2)}.
	\end{array}
	\right.
	\end{split} 
\end{eqnarray*}
Under Scenario 3, we have the true ODR is still linear
while the true ODRs for Scenario 4 and 5 are non-linear due to their $C^Y(\cdot)$ involving covariates-surrogacy interaction. 
Table \ref{table:3} lists the true value $V(\beta_0)$ for each scenario. 

We apply the proposed GEAR with the tensor-product B-splines for Scenario 3-5, respectively. Specifically, we first model $\mu_U(m,x)$ with the tensor-product B-splines of $\{X_U,M_U\}$ in the auxiliary sample. The degree and knots for the B-splines are selected based on five-fold cross validation to minimize the least square error of the linear regression. Then, we search the GEAR within the class of $\mathbb{I}\{\phi_X(X_E)^\top  \beta > 0\}$, where $\phi_X(\cdot)$ is the polynomial basis with degree=2. Here, the augmented term is fitted by a linear regression of $\widehat{\mu}_U(M_{E},X_{E})$ on $\{A_E,\phi_X(X_E)\}$. 
We name the above procedure as `GEAR-Bspline'. For comparison, we also apply the linear procedure described in Section \ref{value_infer} as `GEAR-linear' without taking any basis. One may note both procedures model $\mu_U(m,x)$ incorrectly. Reported in Table \ref{table:3} are the empirical results under GEAR-Bspline and GEAR-linear aggregated over 500 simulations.


\begin{table}[!t]
\caption{Empirical results 
under the GEAR for Scenario 3-5.}\label{table:3} 
\begin{center}
\begin{small}
\begin{sc}
\begin{tabular}{cccc|ccc}
\toprule
		&\multicolumn{3}{c|}{GEAR-linear} &\multicolumn{3}{c}{GEAR-Bspline} \\
		\midrule 
		$N_E=$& $200$ & $400$&$800$& $200$ & $400$&$800$\\
		\midrule 
		\midrule 
		S3&  $V(\beta_0)$&=&1.20\\
		\midrule 
 $\widehat{V}_{AIP}(\widehat{\beta}^{G})$&1.25&1.22 &1.22&1.26 &1.23&1.22 \\
\midrule 
$SE\{\widehat{V}_{AIP}\}$&0.02 &0.01&0.01& 0.02& 0.01 & 0.01 \\
\midrule 
 $\Mean\{\widehat{\sigma}_{AIP}\}$&0.02&0.01&0.01&0.02&0.01&0.01\\
\midrule 
 $V(\widehat{\beta}^G)$ &1.18&1.19&1.19&1.16&1.18&1.18\\
\midrule 
CP (\%) &95.2 &96.0 &92.6 &94.0 &95.4 &94.4 \\
\midrule 
\midrule 
		S4&  $V(\beta_0)$&=&2.59\\
\midrule 
$\widehat{V}_{AIP}(\widehat{\beta}^{G})$&2.37&2.34&2.34&2.55&2.51&2.49 \\
\midrule 
$SE\{\widehat{V}_{AIP}\}$&0.02&0.01&0.01& 0.02& 0.01 & 0.01\\
\midrule 
 $\Mean\{\widehat{\sigma}_{AIP}\}$&0.02&0.01&0.01&0.02&0.01&0.01\\
\midrule 
 $V(\widehat{\beta}^G)$ &2.32&2.32&2.33&2.41& 2.43&2.44\\
\midrule 
CP (\%) &77.6 &66.2 &55.2 &94.6 &92.0 &90.0 \\
\midrule 
\midrule 
		S5&  $V(\beta_0)$&=&3.03\\
\midrule 
$\widehat{V}_{AIP}(\widehat{\beta}^{G})$&2.44&2.40&2.40&3.00&2.97&2.93 \\
\midrule 
$SE\{\widehat{V}_{AIP}\}$&0.02&0.01&0.01& 0.02& 0.01 & 0.01 \\
\midrule 
 $\Mean\{\widehat{\sigma}_{AIP}\}$&0.02&0.01&0.01&0.02&0.01&0.01\\
\midrule 
 $V(\widehat{\beta}^G)$ &2.30&2.32&2.32&2.72& 2.77&2.79\\
\midrule 
CP (\%) &31.6 &17.4 &11.8 &96.0 &92.4 &87.8\\
\bottomrule
\end{tabular}
\end{sc}
\end{small}
\end{center} 
\end{table}

It can be seen from Table \ref{table:3} that the GEAR-Bspline procedure performs reasonably better than the linear procedure under non-linear decision rules. Specifically, in Scenario 3 with only the baseline function $H^Y(\cdot)$ non-linear in $X$, GEAR-linear performs comparable to GEAR-Bspline, as the linear model can well approximate the non-linear baseline function. In Scenario 4 and 5 with more complex non-linear function $C^Y(\cdot)$, GEAR-Bspline outperforms GEAR-linear in terms of smaller bias and higher empirical coverage probabilities of the 95\% CI. For example, GEAR-Bspline achieves $V(\widehat{\beta}^G)=2.43$ in Scenario 4 ($V( {\beta}_0)=2.59$) with coverage probability 92.0\% and $V(\widehat{\beta}^G)=2.77$ in Scenario 5 ($V( {\beta}_0)=3.03$) with coverage probability 92.4\% when $N_E=N_U$, while GEAR-linear can hardly maintain an empirical coverage probability over one third in Scenario 5 due to the severe model misspecification.
Note that because of the interaction between $X$ and $M$ in $C^Y(\cdot)$, the model assumption is still mildly violated even applying the GEAR-Bspline method. Thus, the empirical coverage probabilities of the 95\% CI decreases as the sample size $N_E$ increases.


\section{Real Data Analysis}\label{:sec5}
In this section, we illustrate our proposed method by application to the AIDS Clinical Trials Group Protocol 175 (ACTG 175) data. 
There are 1046 HIV-infected subjects enrolled in ACTG 175, who were randomized to two competitive antiretroviral regimens in equal proportions \citep{hammer1996trial}: zidovudine (ZDV) + zalcitabine (ddC), and ZDV+didanosine (ddI). Denote `ZDV+ddC' as treatment 0, versus `ZDV+ddI' as treatment 1. The long-term outcome of interest ($Y$) is the mean CD4 count (cells/mm3) at 96 $\pm$ 5 weeks. A higher CD4 count usually indicates a stronger immune system. However, about one-third of the patients who received treatment 0 or 1 have missing long-term outcome, which form the experimental sample of interest. 
The rest dataset is used as the auxiliary sample. 


In the experimental sample ($N_E=376$), 187 patients were randomized to treatment 0 and 189 patients to treatment 1, with the propensity score as constant $\pi (x)\equiv 0.503$. The auxiliary sample consists of $N_U=670$ subjects with observed long-term outcome. 
We consider $r=12$ baseline covariates used in \citep{tsiatis2008covariate}: 1) four continuous variables: age (years), weight (kg), CD4 count (cells/mm3) at baseline, and CD8 count (cells/mm3) at baseline; 2) eight categorical variables: hemophilia, homosexual activity, history of intravenous drug use, Karnofsky score (scale of 0-100), race (0=white, 1=non-white), gender (0=female), antiretroviral history (0=naive, 1=experienced), and symptomatic status (0=asymptomatic). Intermediate outcomes contain CD4 count at 20 $\pm$ 5 weeks and CD8 count at 20 $\pm$ 5 weeks. It can be shown in the auxiliary data that intermediate outcomes are highly related to the long-term outcome via a linear regression of $Y_U$ on $\{X_U,M_U\}$.

We apply our proposed `GEAR-linear' and `GEAR-Bspline' described in Section \ref{misspe} to the ACTG 175 data, respectively. Here, to avoid the curse of high dimensionality, we only take the polynomial basis on the continuous variables with degree as 2. 
Reported in Table \ref{table:6} are the estimated mean outcome for each treatment as $\widehat{V}_{AIP}(0)$ and $\widehat{V}_{AIP}(1)$, the estimated value $\widehat{V}_{AIP}(\widehat{\beta}^G)$ with its estimated standard deviation $ \widehat{\sigma}_{AIP} $, the 95\% CI for the estimated value
, and the number of assignments for each treatment.

\begin{table}[!t]
\caption{Comparison results for ACTG 175 data.}\label{table:6}
\begin{center}
\begin{small}
\begin{sc}
\begin{tabular}{c|cc}
\toprule
		&Linear &B-spline \\
				\midrule 
$\widehat{V}_{AIP}(0)$ &327.8&325.5\\
 \midrule
$\widehat{V}_{AIP}(1)$ &334.0&329.0\\
 \midrule
$\widehat{V}_{AIP}(\widehat{\beta}^G)$  [SD] & 351.4 [10.2]&346.4 [9.6]\\
 \midrule
95\% CI for $\widehat{V}_{AIP}(\widehat{\beta}^G)$ & (331.4, 371.3)&(327.7, 365.1)\\
 \midrule
 Assign to `ZDV+ddC'&185&184\\
  \midrule
   Assign to `ZDV+ddI'&191&192\\
   \bottomrule
\end{tabular}
\end{sc}
\end{small}
\end{center} 
\end{table}

It is clear that the proposed GEAR estimation procedure with the B-spline performs reasonably better than the linear procedure. Next, we focus on the results obtained from the GEAR-Bspline method in the experimental sample of interest. 
Our proposed GEAR-Bspline method achieves a value of 346.4 with a smaller standard deviation as 9.6 comparing to GEAR-linear (10.2) in the experimental sample. The GEAR with B-spline assigns 192 patients to `ZDV+ddI' and 184 patients to `ZDV+ddC', which is consistent with the competitive nature of these two treatments. 



\section{Discussion}\label{sec:6}
In this paper, we proposed a new personalized optimal decision policy when the long-term outcome of interest cannot be observed. 
Theoretically, we gave the cubic convergence rate of our proposed GEAR, and derived the consistency and asymptotical distributions of the value function under the GEAR. Empirically, we validated our method, and examined the sensitivity of our proposed GEAR when the model is misspecified or when assumptions are violated. 

There are several other possible extensions we may consider in future work. First, we only consider two treatment options in this paper, while in applications it is common to have more than two options for decision making. Thus, a more general method with multiple treatments or even continuous decision marking is desirable. Second, we can extend our work to dynamic decision making, 
where the ultimate outcome of interest cannot be observed in the experimental sample but can be found in some auxiliary dataset.

 \newpage
\bibliographystyle{agsm}
\bibliography{mycite}

\newpage
\appendix
 
 This appendix is organized as follows. In Section \ref{ipw_sec}, we provide the inverse propensity-score weighted (IPW) value estimator and its related theories as a middle step. In Section \ref{proofs}, we give technical proofs for all the established theoretical results. Section \ref{add_simu} presents additional simulation results for sensitivity studies under scenarios with model misspecification and assumptions violation.

\appendix

\section{Inverse Propensity-Score Weighted Estimator}\label{ipw_sec}

\subsection{IPW Estimator for the Long-term Outcome}
 
According to Lemma \ref{lem1} and the law of large number, the value function $V(\beta)$ can be consistently estimated by
\begin{eqnarray*}
V_n(\beta)=   {1\over N_E}\sum_{i=1}^{N_E} {\mathbb{I}\{A_{E,i}=d(X_{E,i};\beta)\}{\mu}_U(M_{E,i},X_{E,i})\over{A_{E,i} \pi (X_{E,i})+(1-A_{E,i})\{1-\pi(X_{E,i})\}}}  . 
\end{eqnarray*}

We posit parametric models for $\pi (x)\equiv \pi (x;\gamma)$ and $\mu_U(m,x)\equiv \mu_U(m,x;\lambda)$ with the true model parameter $\gamma$ and $\lambda$, respectively. 
Then the above $V_n(\beta)$ can be rewritten as the model-based form,
\begin{eqnarray*}
V^{\star}_n(\beta)=   {1\over N_E}\sum_{i=1}^{N_E} {\mathbb{I}\{A_{E,i}=d(X_{E,i};\beta)\}{\mu}_U(M_{E,i},X_{E,i};\lambda)\over{A_{E,i} \pi (X_{E,i};\gamma)+(1-A_{E,i})\{1-\pi(X_{E,i};\gamma)\}}} ,
\end{eqnarray*}
where $\pi (x;\gamma)$ can be estimated in the experimental sample, denoted as $\pi(x;\widehat{\gamma})$, and $\mu_U(m,x;\lambda)$ can be estimated in the auxiliary sample, denoted as $\mu_U(m,x;\widehat{\lambda})$. 
Then, by replacing the implicit functions in $V^{\star}_n(\beta)$ with their parametric estimators, it is straightforward to give the following IPW estimator for the value function $V(\beta)$,
\begin{eqnarray}\label{ipw}
\widehat{V}(\beta) = {1\over N_E}\sum_{i=1}^{N_E} {\mathbb{I}\{A_{E,i}=d(X_{E,i},\beta)\} \mu_U(M_{E,i},X_{E,i};\widehat{\lambda})\over{A_{E,i} \pi(X_{E,i};\widehat{\gamma})+(1-A_{E,i})\{1-\pi(X_{E,i};\widehat{\gamma})\}}}.
\end{eqnarray}   

Define $\widehat{\beta}=\underset{\beta}{\argmax} \widehat{V}(\beta)$ with subject to $||\beta||_2=1$ for identifiability purpose, with the corresponding estimated value function $\widehat{V}(\widehat{\beta})$.

\subsection{Theoretical Results of the IPW Estimator}

First, we establish some theoretical results for the IPW estimator as a middle step to prove the results for the AIPW estimator. Here, we use $\phi_X(X)$ and $\phi_M(M)$ to represent appropriate basis functions for $X$ and $M$, respectively. The following theorem gives the consistency result of our IPW estimator for the value function to the true. The proof is provided in Section \ref{ipw_proof}.

\begin{thm}{(Consistency)}\label{thm1}
When (A1)-(A9) and (A11) hold, given $\forall \beta$, we have 
\begin{eqnarray*}
\widehat{V}(\beta)=V(\beta)+o_p(1).
\end{eqnarray*}
\end{thm}

\bigskip

Next, we establish the asymptotic normality of $\sqrt{N_E}\big\{\widehat{V}(\widehat{\beta})-V(\beta_0)\big\}$ through the following lemma that states the estimator $\widehat{\beta}$ has a cubic rate towards the true $\beta_0$. The proof is provided in Section \ref{ipw_proof2}.
\begin{lemma}\label{lem3ipw}
Under (A1)-(A11), we have
\begin{eqnarray}\label{cubic}
N_E^{1/3}||\widehat{\beta}-\beta_0||_2 = O_p(1),
\end{eqnarray}
where $||\cdot||_2$ is the $L_2$ norm.
\end{lemma}


We next show the asymptotic distribution of $ \widehat{V}(\widehat{\beta}) $ as follows. The proof is provided in Section \ref{ipw_proof3}.
\begin{thm}{(Asymptotic Distribution)}\label{thm2}
When (A1)-(A11) are satisfied, we have 
\begin{eqnarray}\label{asyipw}
\sqrt{N_E}\big\{\widehat{V}(\widehat{\beta})-V(\beta_0)\big\}\overset{\mathcal{D}}{\longrightarrow} N(0,\sigma_{IPW}^2),
\end{eqnarray}
where $\sigma_{IPW}^2=t\sigma_{U}^2+\sigma_{E,I}^2$, and $\sigma_{U}^2=\Mean[\{\xi_{i}^{({U})}\}^2]$ and $\sigma_{E,I}^2=\Mean[\{\xi_{i}^{(E,I)}\}^2]$.

Here, $ \xi_{i}^{(U)}\equiv G_2^\top H_2^{-1}\begin{bmatrix}
   			\phi_X(X_{U,i})\\
   			\phi_M(M_{U,i})
		\end{bmatrix} \{Y_{U,i}-{\mu}_U(M_{U,i},X_{U,i};\lambda)\} $ is the I.I.D. term in the auxiliary sample, and $\xi_{i}^{(E,I)}\equiv G_1^\top  H_1^{-1} \phi_X(X_{E,i})\{A_{E,i}-\pi (X_{E,i};\gamma)\}+ {\mathbb{I}\{A_{E,i}=d(X_{E,i};\beta_0)\} \mu_U(M_{E,i},X_{E,i}; {\lambda})\over{A_{E,i} \pi(X_{E,i}; {\gamma})+(1-A_{E,i})\{1-\pi(X_{E,i}; {\gamma})\}}}-V(\beta_0) $ is the I.I.D. term in the experimental sample.
\end{thm}

\section{Technical Proofs}\label{proofs}
 		
\subsection{Proof of Lemma \ref{lem1}}
For any decision rule $d(\cdot;\beta)$, we will show that $V_n(\beta)$ is a consistent estimator of the value function $V(\beta)$ under (A1)-(A5). 

\smallskip

(A.) First, we rewrite the expectation of the I.I.D. summation term of $V_n(\beta)$ using the law of iterated expectation with (A4) and (A5).

(a1.) Taking the iterated expectation on $\{X_E, M_E\}$, we have 
\begin{eqnarray*}
\begin{split}
\Mean \{V_n(\beta)\}=&\Mean \Bigg[{\mathbb{I}\{A_E=d(X_E;\beta)\}\over{A_E\pi(X_E)+(1-A_E)\{1-\pi(X_E)\}}}\mu_U(M_E,X_E)\Bigg]\\
=&\Mean \Bigg[ \Mean \Bigg\{{\mathbb{I}\{A_E=d(X_E;\beta)\}\over{A_E\pi(X_E)+(1-A_E)\{1-\pi(X_E)\}}}\mu_U(M_E,X_E)\Bigg|X_E,M_E\Bigg\}\Bigg]\\
=&\Mean \Bigg[\mu_U(M_E,X_E) \Mean \Bigg\{{\mathbb{I}\{A_E=d(X_E;\beta)\}\over{A_E\pi(X_E)+(1-A_E)\{1-\pi(X_E)\}}}\Bigg|X_E,M_E\Bigg\}\Bigg].\\
\end{split}
\end{eqnarray*}

(a2.) By Corollary \ref{coro1} that $\mu_U(M_E,X_E)=\Mean  \{Y_U|X_E,M_E \}=\Mean  \{Y_E|X_E,M_E \}$, thus
\begin{eqnarray*}
\Mean \{V_n(\beta)\}=\Mean \Bigg[\Mean  \{Y_E|X_E,M_E \} \Mean \Bigg\{{\mathbb{I}\{A_E=d(X_E;\beta)\}\over{A_E\pi(X_E)+(1-A_E)\{1-\pi(X_E)\}}}\Bigg|X_E,M_E\Bigg\}\Bigg].
\end{eqnarray*}

(a3.) By (A5) that $Y_E \independent A_E\mid X_E,M_E$, we have
\begin{eqnarray*}
\Mean \{V_n(\beta)\}=\Mean \Bigg[\Mean \Bigg\{{\mathbb{I}\{A_E=d(X_E;\beta)\}\over{A_E\pi(X_E)+(1-A_E)\{1-\pi(X_E)\}}}Y_E\Bigg|X_E,M_E\Bigg\}\Bigg]. 
\end{eqnarray*}

(a4.) From the inverse of the law of iterated expectation, then 
\begin{eqnarray*}
\Mean \{V_n(\beta)\}=\Mean \Bigg[{\mathbb{I}\{A_E=d(X_E;\beta)\}\over{A_E\pi(X_E)+(1-A_E)\{1-\pi(X_E)\}}}Y_E\Bigg].
\end{eqnarray*}

\smallskip

(B.) Next, we proof above $\Mean \{V_n(\beta)\}$ is a consistent estimator of the value function $V(\beta)$ under (A1)-(A3).

(b1.) Taking the iterated expectation on $\{X_E\}$, we have 
\begin{eqnarray*}
\Mean \{V_n(\beta)\}=\Mean \Bigg[\Mean \Bigg\{{\mathbb{I}\{A_E=d(X_E;\beta)\}\over{A_E\pi(X_E)+(1-A_E)\{1-\pi(X_E)\}}}Y_E\Bigg|X_E\Bigg\}\Bigg].\\
\end{eqnarray*}

(b2.) Use the fact that $\mathbb{I}\{A_E=d(X_E;\beta)\}=Ad(X_E;\beta)+(1-A_E)\{1-d(X_E;\beta)\}$, thus
\begin{eqnarray*}
\begin{split}
\Mean \{V_n(\beta)\}=&\Mean \Bigg[\Mean \Bigg\{{{AY_E }\over{A_E\pi(X_E)+(1-A_E)\{1-\pi(X_E)\}}}\Bigg|X_E\Bigg\}d(X_E;\beta)\\
&+\Mean \Bigg\{{{(1-A)Y_E }\over{A_E\pi(X_E)+(1-A_E)\{1-\pi(X_E)\}}}\Bigg|X_E\Bigg\}\{1-d(X_E;\beta)\}\Bigg].\\
\end{split}
\end{eqnarray*}

(b3.) By (A1) that $Y_E= A Y_E^{\star}(1)  + (1-A)Y_E^{\star}(0)$, and the fact $AA=A$, 
\begin{eqnarray*}
\begin{split}
\Mean \{V_n(\beta)\}=&\Mean \Bigg[\Mean \Bigg\{{{A\{A Y_E^{\star}(1)  + (1-A)Y_E^{\star}(0)\} }\over{A_E\pi(X_E)+(1-A_E)\{1-\pi(X_E)\}}}\Bigg|X_E\Bigg\}d(X_E;\beta)\\
&+\Mean \Bigg\{{{(1-A_E)\{A Y_E^{\star}(1)  + (1-A)Y_E^{\star}(0)\} }\over{A_E\pi(X_E)+(1-A_E)\{1-\pi(X_E)\}}}\Bigg|X_E\Bigg\}\{1-d(X_E;\beta)\}\Bigg]\\
=&\Mean \Bigg[\Mean \Bigg\{{{A Y_E^{\star}(1)  }\over{A_E\pi(X_E)+(1-A_E)\{1-\pi(X_E)\}}}\Bigg|X_E\Bigg\}d(X_E;\beta)\\
&+\Mean \Bigg\{{{(1-A)Y_E^{\star}(0) }\over{A_E\pi(X_E)+(1-A_E)\{1-\pi(X_E)\}}}\Bigg|X_E\Bigg\}\{1-d(X_E;\beta)\}\Bigg].\\
\end{split}
\end{eqnarray*}

(b4.) Applying (A2) that $\{Y_E^*(0),Y_E^*(1)\}\independent A\mid X_E$, we have 

\begin{eqnarray*}
\begin{split}
\Mean \{V_n(\beta)\}=&\Mean \Bigg[\Mean \{Y_E^{\star}(1) |X_E \} \Mean \Bigg\{{{A}\over{A_E\pi(X_E)+(1-A_E)\{1-\pi(X_E)\}}}\Bigg|X_E\Bigg\}d(X_E;\beta)\\
&+\Mean \{Y_E^{\star}(0) |X_E \} \Mean \Bigg\{{{(1-A)}\over{A_E\pi(X_E)+(1-A_E)\{1-\pi(X_E)\}}}\Bigg|X_E\Bigg\}\{1-d(X_E;\beta)\}\Bigg]\\
\end{split}
\end{eqnarray*}
\begin{eqnarray*}
\begin{split}
=&\Mean \Bigg[\Mean  \{Y_E^{\star}(1) |X_E \}\Mean\Bigg\{{1\over{\pi(X_E)}}P(A_E=1|X_E)\Bigg\}d(X_E;\beta)\\
&+\Mean \{Y_E^{\star}(0) |X_E \} \Mean \Bigg\{{{1}\over{1-\pi(X_E)}}P(A_E=0|X_E)\Bigg\}\{1-d(X_E;\beta)\}\Bigg].\\
\end{split}
\end{eqnarray*}

(b5.) Based on (A3) that $0<\pi(x)<1$ for all $x \in \mathbb{X}_E$, as well as the inverse of the iterated expectation, we finally show that
\begin{eqnarray*}
\begin{split}
\Mean \{V_n(\beta)\}=&\Mean \Bigg[\Mean  \{Y_E^{\star}(1) |X_E \}d(X_E;\beta)+\Mean \{Y_E^{\star}(0) |X_E \}\{1-d(X_E;\beta)\}\Bigg]\\
=&\Mean [Y_E^*(0) \{1-d(X_E;\beta)\}+Y_E^*(1) d(X_E;\beta)]=V(\beta).  \quad \square
\end{split}
\end{eqnarray*}

\subsection{Proof of Lemma \ref{lem2}}
Lemma \ref{lem2} can be easily shown through the technic of the law of iterated expectation with (A4) and (A5). 

(A.) By taking the iterated expectation on $\{M_E\}$, we have
\begin{eqnarray*}
\Mean_{Y_E|X_E}\{Y_E|A_E=d(X_E;\beta),X_E\}=\Mean_{M_E|X_E}[\Mean_{Y_E}\{Y_E|A_E=d(X_E;\beta),X_E,M_E\}].
\end{eqnarray*}

(B.) By (A5) that $Y_E \independent A_E\mid X_E,M_E$, we have
\begin{eqnarray*}
\Mean_{M_E|X_E}[\Mean_{Y_E}\{Y_E|A_E=d(X_E;\beta),X_E,M_E\}]=\Mean_{M_E|X_E}[\Mean_{Y_E}\{Y_E|M_E,X_E\}|A_E=d(X_E;\beta),X_E].
\end{eqnarray*}

(C.) By Corollary \ref{coro1} that $\mu_U(M_E,X_E)=\Mean  \{Y_U|X_E,M_E \}=\Mean  \{Y_E|X_E,M_E \}$, thus
\begin{eqnarray*}
\Mean_{M_E|X_E}[\Mean_{Y_E}\{Y_E|M_E,X_E\}|A_E=d(X_E;\beta),X_E]=\Mean_{M_E|X_E}[\mu_U(M_E,X_E)|A_E=d(X_E;\beta),X_E].\quad \square 
\end{eqnarray*} 
 
\subsection{Proof of Theorem \ref{thm1}}\label{ipw_proof}
To show that $\widehat{V}(\beta)=V(\beta)+o_p(1)$, it is sufficient to show that $V^{\star}_n(\beta)=V(\beta)+o_p(1)$ and $\widehat{V}(\beta)=V^{\star}_n(\beta)+O_E(N_E^{-{1\over2}})$.

\bigskip

(A.) First, given a decision rule $d(\cdot;\beta)$, by the Weak Law of Large Number, we have 
\begin{eqnarray*}
\begin{split}
V^{\star}_n(\beta)=   {1\over N_E}\sum_{i=1}^{N_E} & {\mathbb{I}\{A_{E,i}=d(X_{E,i};\beta)\}\over{A_{E,i} \pi (X_{E,i};\gamma)+(1-A_{E,i})\{1-\pi(X_{E,i};\gamma)\}}} {\mu}_U(M_{E,i},X_{E,i};\lambda)\\&\underset{p}{\longrightarrow} \Mean \Bigg[{\mathbb{I}\{A_E=d(X_E;\beta)\}\over{A_E\pi(X_E;\gamma)+(1-A_E)\{1-\pi(X_E;\gamma)\}}}\mu_U(M_E,X_E;\lambda)\Bigg]= V(\beta).
\end{split}
\end{eqnarray*}
That is, $V^{\star}_n(\beta)=V(\beta)+o_p(1)$. 

\bigskip

(B.) Next, we show the following is $O_E(N_E^{-{1\over2}})$, 
\begin{eqnarray}\label{funf}
\begin{split}
\widehat{V}(\beta)-V^{\star}_n(\beta)= &{1\over N_E}\sum_{i=1}^{N_E}\Bigg[\underbrace{ {\mathbb{I}\{A_{E,i}=d(X_{E,i};\beta)\}\over{A_{E,i} \pi(X_{E,i};\widehat{\gamma})+(1-A_{E,i})\{1-\pi(X_{E,i};\widehat{\gamma})\}}} \mu_U(M_{E,i},X_{E,i};\widehat{\lambda})}_{f_i(\widehat{\gamma},\widehat{\lambda};\beta)}\\
&-\underbrace{{\mathbb{I}\{A_{E,i}=d(X_{E,i};\beta)\}\over{A_{E,i} \pi(X_{E,i};\gamma)+(1-A_{E,i})\{1-\pi(X_{E,i};\gamma)\}}} {\mu}_U(M_{E,i},X_{E,i};\lambda)}_{f_i( {\gamma}, {\lambda};\beta)}\Bigg].\\
\end{split}
\end{eqnarray}

(b1.) By (A8), with appropriate parametric model for $ \mu_U(m,x;\lambda)$, we can present the estimator $\widehat{\lambda}$ in the auxiliary sample as
\begin{eqnarray}\label{lambda}
\sqrt{N_U}(\widehat{\lambda}-\lambda)=H_2^{-1} {1\over \sqrt{N_U}}\sum_{i=1}^{N_U} \begin{bmatrix}
   			\phi_X(X_{U,i})\\
   			\phi_M(M_{U,i})
		\end{bmatrix} \{Y_{U,i}-{\mu}_U(M_{U,i},X_{U,i};\lambda)\}+o_p(1),
\end{eqnarray}
where $H_2\equiv \underset{N_U\to +\infty}{\lim} {1\over N_U}\sum_{i=1}^{N_U} \begin{bmatrix}
   			\phi_X(X_{U,i})\\
   			\phi_M(M_{U,i})
		\end{bmatrix}\{\partial {\mu}_U(M_{U,i},X_{U,i};\lambda)/\partial\lambda\}^\top $.
		
Similarly, according to (A11), for the IPW estimator, we can present the estimator $\widehat{\gamma}$ in the experimental sample as
\begin{eqnarray}\label{gamma}
\sqrt{N_E}(\widehat{\gamma}-\gamma)=H_1^{-1} {1\over \sqrt{N_E}}\sum_{i=1}^{N_E} \phi_X(X_{E,i})\{A_{E,i}-\pi (X_{E,i};\gamma)\}+o_p(1),
\end{eqnarray}
where $H_1\equiv \underset{N_E\to +\infty}{\lim} {1\over N_E}\sum_{i=1}^{N_E} \phi_X(X_{E,i}) \{\partial \pi (X_{E,i};\gamma)/\partial\gamma\}^\top $.
		
Take the Taylor Expansion on $f_i(\widehat{\gamma},\widehat{\lambda};\beta)$ at $( {\gamma}, {\lambda})$, we have
\begin{eqnarray}\label{taylor}
\begin{split}
&f_i(\widehat{\gamma},\widehat{\lambda};\beta)-f_i( {\gamma}, {\lambda};\beta)=
\{\partial f_i(\widehat{\gamma},\widehat{\lambda})/\partial\gamma\}(\widehat{\gamma}-{\gamma})+\{\partial f_i(\widehat{\gamma},\widehat{\lambda})/\partial\lambda\}(\widehat{\lambda}-{\lambda})\\
=&{\mathbb{I}\{A_{E,i}=d(X_{E,i};\beta)\} {\mu}_U(M_{E,i},X_{E,i};\lambda)\over{[A_{E,i} \pi(X_{E,i};\gamma) +(1-A_{E,i})\{1-\pi(X_{E,i};\gamma)\}]^2}}(1-2A_{E,i})\{\partial \pi (X_{E,i};\gamma)/\partial\gamma\}(\widehat{\gamma}-{\gamma})\\
&+{\mathbb{I}\{A_{E,i}=d(X_{E,i};\beta)\}\over{A_{E,i} \pi(X_{E,i};\gamma)+(1-A_{E,i})\{1-\pi(X_{E,i};\gamma)\}}} \{\partial\mu_U(M_{E,i},X_{E,i};\lambda)/\partial \lambda\}(\widehat{\lambda}-{\lambda})+o_p(N_E^{-{1\over2}}).\\
\end{split}
\end{eqnarray}

(b2.) Let 
\begin{eqnarray*}
\xi_{i,1}={\mathbb{I}\{A_{E,i}=d(X_{E,i};\beta)\} {\mu}_U(M_{E,i},X_{E,i};\lambda)\over{[A_{E,i} \pi(X_{E,i};\gamma) +(1-A_{E,i})\{1-\pi(X_{E,i};\gamma)\}]^2}}(1-2A_{E,i})\{\partial \pi (X_{E,i};\gamma)/\partial\gamma\},
\end{eqnarray*}
and 
\begin{eqnarray*}
\xi_{i,2}={\mathbb{I}\{A_{E,i}=d(X_{E,i};\beta)\}\over{A_{E,i} \pi(X_{E,i};\gamma)+(1-A_{E,i})\{1-\pi(X_{E,i};\gamma)\}}} \{\partial\mu_U(M_{E,i},X_{E,i};\lambda)/\partial \lambda\},
\end{eqnarray*}
by (A3) that $0<\pi (x;\gamma)<1$, and (A7) that $\mu_U(m,x;\lambda)$, $\partial \pi (x;\gamma)/\partial\gamma$ and $\partial\mu_U(m,x;\lambda)/\partial \lambda$ are bounded, applying the Weak Law of Large Number, we have
${1\over N_E}\sum_{i=1}^{N_E} \xi_{i,1} \underset{p}{\longrightarrow} B_1< \infty$, and ${1\over N_E}\sum_{i=1}^{N_E} \xi_{i,2} \underset{p}{\longrightarrow} B_2 < \infty$, as $N_E \rightarrow \infty$.

(b3.) By rearranging the equations, we have 
\begin{eqnarray*}
\begin{split}
&\widehat{V}(\beta)-V^{\star}_n(\beta)={1\over N_E}\sum_{i=1}^{N_E} \{f_i(\widehat{\gamma},\widehat{\lambda};\beta)-f_i( {\gamma}, {\lambda};\beta)\}\\
=&(\widehat{\gamma}-{\gamma}) {1\over N_E}\sum_{i=1}^{N_E}  \xi_{i,1} +(\widehat{\lambda}-{\lambda}) {1\over N_E}\sum_{i=1}^{N_E}  \xi_{i,2}+o_p(N_E^{-{1\over2}}),\\
\end{split}
\end{eqnarray*}
where $\widehat{\gamma}-{\gamma}=O_E(N_E^{-{1\over2}}) $, $\widehat{\lambda}-{\lambda}=O_E(N_U^{-{1\over2}})$, ${1\over N_E}\sum_{i=1}^{N_E}  \xi_{i,1}=o_p(1)+B_1$, and ${1\over N_E}\sum_{i=1}^{N_E}  \xi_{i,2}=o_p(1)+B_2$. By (A9) that $t=\sqrt{{N_E\over N_U}}$ with $0<t<+\infty$, thus,
\begin{eqnarray*}
\widehat{V}(\beta)-V^{\star}_n(\beta)= O_E(N_E^{-{1\over2}}).
\end{eqnarray*}

Therefore, $\widehat{V}(\beta)-V(\beta)=\widehat{V}(\beta)-V^{\star}_n(\beta)+V^{\star}_n(\beta)-V(\beta)= O_E(N_E^{-{1\over2}})+o_p(1)=o_p(1)$. $\square$

\subsection{Proof of Lemma \ref{lem3ipw} and Lemma \ref{lem3}}\label{ipw_proof2}
(A.) First, we show that $\widehat{\beta}$ converges in probability to $\beta_0$ as $N_E \to \infty$, by checking three conditions of the Argmax Theorem:

(a1.) By (A10) that the true value function $V(\beta)$ has twice continuously differentiable at an inner point of maximum $\beta_0$.

(a2.) By the conclusion of Theorem \ref{thm1}, $\widehat{V}(\beta)=V(\beta)+o_p(1)$, i.e., for $\forall \beta$
\begin{eqnarray*}
\widehat{V}(\beta)\overset{p}{\longrightarrow} V(\beta),\quad \text{as}\quad N_E \to \infty.
\end{eqnarray*}

(a3.)  Since $\widehat{\beta}=\underset{\beta}{\argmax} \widehat{V}(\beta)$, we have the estimated ODR as $d(X_E;\widehat{\beta})=\mathbb{I}\{\phi_X(X_E)^\top  \widehat{\beta}>0\}$ and the corresponding value function $\widehat{V}(\widehat{\beta})$ such that 
$$
\widehat{V}(\widehat{\beta}) \geq sup_{\beta \in \rm B} \widehat{V}(\beta).
$$

Thus, we have $\widehat{\beta}  \overset{p}{\longrightarrow} \beta_0$ as $N_E \to \infty$.

\bigskip

(B.) Next, we show that the convergence rate of $\widehat{\beta}$ is $N_E^{1/3}$, i.e. $N_E^{1/3}||\widehat{\beta}-\beta_0||_2 = O_p(1)$, where $||\cdot||_2$ is $L_2$ norm, via checking three conditions of the Theorem 14.4: Rate of convergence in \cite{kosorok2008introduction}:

(b1.) For every $\beta$ in a neighborhood of $\beta_0$, i.e. $||\beta- \beta_0||_2<\delta$,  by (A10), we take the second order Taylor expansion of $V(\beta)$ at $\beta=\beta_0$, 
\begin{eqnarray*}
\begin{split}
V(\beta) -V(\beta_0)=&V'(\beta_0)||\beta- \beta_0||_2+{1\over 2}V''(\beta_0)||\beta- \beta_0||_2^2+o\{||\beta- \beta_0||_2^2\}\\
(\text{by }V'(\beta_0)=0)=&{1\over 2}V''(\beta_0)||\beta- \beta_0||_2^2+o\{||\beta- \beta_0||_2^2\}.
\end{split}
\end{eqnarray*}
Since $V''(\beta_0)<0$, there exist $c_1=-{1\over 2}V''(\beta_0)>0$ such that
$V(\beta) -V(\beta_0) \leq c_1||\beta- \beta_0||_2^2$ holds.

(b2.) For all $N_E$ large enough and sufficiently small $\delta$, the centered process $\widehat{V}-V$ satisfies
\begin{eqnarray*}
\begin{split}
&\Mean^{*} \underset{||\beta-\beta_0||_2<\delta}{\sup} \sqrt{N_E} |\widehat{V}(\beta)-V(\beta) -\{\widehat{V}(\beta_0)-V(\beta_0)\}|\\
=&\Mean^{*} \underset{||\beta-\beta_0||_2<\delta}{\sup} \sqrt{N_E} |\widehat{V}(\beta)-V^{\star}_n(\beta)+V^{\star}_n(\beta)-V(\beta) -\{\widehat{V}(\beta_0)-V^{\star}_n(\beta_0)+V^{\star}_n(\beta_0)-V(\beta_0)\}|\\
\end{split}
\end{eqnarray*}
\begin{eqnarray*}
\begin{split}
\leq&\underbrace{\Mean^{*} \underset{||\beta-\beta_0||_2<\delta}{\sup} \sqrt{N_E} \Big|\widehat{V}(\beta)-V^{\star}_n(\beta)-\{\widehat{V}(\beta_0)-V^{\star}_n(\beta_0)\}\Big|}_{\eta_1}\\
&+\underbrace{\Mean^{*} \underset{||\beta-\beta_0||_2<\delta}{\sup} \sqrt{N_E} \Big|V^{\star}_n(\beta)-V(\beta) -\{V^{\star}_n(\beta_0)-V(\beta_0)\}\Big|}_{\eta_2},
\end{split}
\end{eqnarray*}
where $\Mean^{*} $ is the outer expectation.

 We first derive two results (b2.1) and (b2.2) to bound ${\eta_1}$ and ${\eta_2}$, respectively, and then we are able to show the second condition of Theorem 14.4 is satisfied.

(b2.1) First, recalling the definition made in Equation (\ref{funf}) that
\begin{eqnarray*}
f_i( {\gamma}, {\lambda};\beta)={\mathbb{I}\{A_{E,i}=d(X_{E,i};\beta)\}\over{A_{E,i} \pi(X_{E,i};\gamma)+(1-A_{E,i})\{1-\pi(X_{E,i};\gamma)\}}} {\mu}_U(M_{E,i},X_{E,i};\lambda),
\end{eqnarray*}
with the fact that
\begin{eqnarray}\label{transrule}
\begin{split}
&\mathbb{I}\{A_{E,i}=d(X_{E,i};\beta)\}-\mathbb{I}\{A_{E,i}=d(X_{E,i};\beta_0)\}\\
=&A_{E,i} \mathbb{I}\{\phi_X(X_{E,i})^\top \beta>0\}+(1-A_{E,i})\Big[1- \mathbb{I}\{\phi_X(X_{E,i})^\top \beta>0\}\Big]\\
&-A_{E,i} \mathbb{I}\{\phi_X(X_{E,i})^\top \beta_0>0\}-(1-A_{E,i})\Big[1- \mathbb{I}\{\phi_X(X_{E,i})^\top \beta_0>0\}\Big]\\
=&(2A_{E,i}-1)\Big[\mathbb{I}\{\phi_X(X_{E,i})^\top \beta>0\}-\mathbb{I}\{\phi_X(X_{E,i})^\top \beta_0>0\}\Big],
\end{split}
\end{eqnarray}
we have,
\begin{eqnarray}\label{diff}
\begin{split}
&V^{\star}_n(\beta)-V^{\star}_n(\beta_0)={1\over N_E}\sum_{i=1}^{N_E}\{ f_i( {\gamma}, {\lambda};\beta)-f_i( {\gamma}, {\lambda};\beta_0)\}\\
=&{1\over N_E}\sum_{i=1}^{N_E} {{\mu_U(M_{E,i},X_{E,i};\lambda)\big[\mathbb{I}\{A_{E,i}=d(X_{E,i};\beta)\}-\mathbb{I}\{A_{E,i}=d(X_{E,i};\beta_0)\}\big]}\over {A_{E,i} \pi(X_{E,i};\gamma)+(1-A_{E,i})\{1-\pi(X_{E,i};\gamma)\}}}\\
=&{1\over N_E}\sum_{i=1}^{N_E} {{\mu_U(M_{E,i},X_{E,i};\lambda)(2A_{E,i}-1)\big[\mathbb{I}\{\phi_X(X_{E,i})^\top \beta>0\}-\mathbb{I}\{\phi_X(X_{E,i})^\top \beta_0>0\}\big]}\over {A_{E,i} \pi(X_{E,i};\gamma)+(1-A_{E,i})\{1-\pi(X_{E,i};\gamma)\}}}.\\
\end{split}
\end{eqnarray}

Then, we define a class of function
$$
\mathcal{F}_\beta^{1} (x,a,m)=\Bigg\{{{\mu_U(m,x;\lambda)(2a-1)\big[\mathbb{I}\{\phi_X(x)^\top \beta>0\}-\mathbb{I}\{\phi_X(x)^\top \beta_0>0\}\big]}\over {a \pi (x;\gamma)+(1-a )\{1-\pi(x;\gamma)\}}} : ||\beta-\beta_0||_2<\delta\Bigg\}.
$$

Let $M_1=\sup \big|{{\mu_U(m,x;\lambda)(2a-1)}\over {a\pi (x;\gamma)+(1-a )\{1-\pi(x;\gamma)\}}}\big|$, by (A7) that $\mu_U(m,x;\lambda)$ and $\pi(x;\gamma)$ both bounded, we have $M_1<\infty$. Then, we define the envelope of $\mathcal{F}_\beta^{1} $ as $F_1=M_1 \mathbb{I}\{1-\delta\leq \phi_X(x)^\top \beta_0 \leq1+\delta\}$; by by (A6) that the density function of covariate $f_X(x)$ is bounded away from 0 and $\infty$, thus,
\begin{eqnarray*}
||F_1||_{P,2}=M_1\sqrt{Pr\{1-\delta\leq \phi_X(x)^\top \beta_0 \leq1+\delta\}}=M_1\sqrt{f_X(\beta_0)\cdot 2\delta}= M_1\sqrt{2f_X(\beta_0)} \delta^{1\over2}<\infty.
\end{eqnarray*}
Since $\mathcal{F}_\beta^{1} $ is an indicate function, by the conclusion of the Lemma 2.6.15 and Lemma 2.6.18 (iii) in \cite{wellner2013weak}, $\mathcal{F}_\beta^{1} $ is a VC (and hence Donsker) class of functions. Thus, the entropy of the class function $\mathcal{F}_\beta^{1} $ denoted as $J_{[]}^{*}(1,\mathcal{F}^{1} )$ is finite, i.e., $J_{[]}^{*}(1,\mathcal{F}^{1} )<\infty$. 

Next, we consider the following empirical process indexed by $\beta$,
\begin{eqnarray*}
\mathbb{G}_n \mathcal{F}_\beta^{1} ={1\over \sqrt{N_E}}\sum_{i=1}^{N_E} \Big\{\mathcal{F}_\beta^{1}  (X_{E,i},A_{E,i},M_{E,i})-\Mean \mathcal{F}_\beta^{1}  (X_{E,i},A_{ i},M_{E,i})\Big\}.
\end{eqnarray*}
Note that $\mathbb{G}_n \mathcal{F}_\beta^{1} =\sqrt{N_E}\big[V^{\star}_n(\beta)-V^{\star}_n(\beta_0)-\{V (\beta)-V (\beta_0)\}\big]$ by Equation (\ref{diff}). Therefore, by applying Theorem 11.2 in \cite{kosorok2008introduction}, we have,
\begin{eqnarray*}
\begin{split}
{\eta_2}=&\Mean^{*} \underset{||\beta-\beta_0||_2<\delta}{\sup} \sqrt{N_E} \Big|V^{\star}_n(\beta)-V(\beta) -\{V^{\star}_n(\beta_0)-V(\beta_0)\}\Big|\\
=&\Mean^{*} \underset{||\beta-\beta_0||_2<\delta}{\sup} \Big|\mathbb{G}_n \mathcal{F}_\beta^{1}  \Big|\leq c_1J_{[]}^{*}(1,\mathcal{F}^{1} ) ||F_1||_{P,2}=c_1J_{[]}^{*}(1,\mathcal{F}^{1} )M_1\sqrt{2f_X(\beta_0)} \delta^{1\over2},
\end{split}
\end{eqnarray*}
where and $c_1$ is a finite constant. 

Let $C_1^{*} \equiv c_1J_{[]}^{*}(1,\mathcal{F}^{1} ) M_1\sqrt{2f_X(\beta_0)}$, since $J_{[]}^{*}(1,\mathcal{F}^{1} )$, $M_1$, and $f_X(\cdot)$ are bounded, we have $C_1^{*}<\infty$, i.e.,
\begin{eqnarray}\label{b1}
{\eta_2}\leq C_1^{*} \delta^{1\over2}.
\end{eqnarray}

\bigskip

(b2.2). First, we rewrite the form of $\widehat{V}(\beta)-V^{\star}_n(\beta)-\{\widehat{V}(\beta_0)-V^{\star}_n(\beta_0)\}$ by Equation (\ref{funf}) and Equation (\ref{transrule}) as
\begin{eqnarray*}
\begin{split}
&\widehat{V}(\beta)-V^{\star}_n(\beta)-\{\widehat{V}(\beta_0)-V^{\star}_n(\beta_0)\}={1\over N_E}\sum_{i=1}^{N_E}\Big[f_i(\widehat{\gamma},\widehat{\lambda};\beta)-f_i( {\gamma}, {\lambda};\beta)-\{f_i(\widehat{\gamma},\widehat{\lambda};\beta_0)-f_i( {\gamma}, {\lambda};\beta_0)\}\Big]
\end{split}
\end{eqnarray*}
\begin{eqnarray*}
\begin{split}
=&{1\over N_E}\sum_{i=1}^{N_E}\Bigg\{(2A_{E,i}-1)\Big[\mathbb{I}\{\phi_X(X_{E,i})^\top \beta>0\}-\mathbb{I}\{\phi_X(X_{E,i})^\top \beta_0>0\}\Big] \\
&\times\Big[{\mu_U(M_{E,i},X_{E,i};\widehat{\lambda})\over{A_{E,i} \pi(X_{E,i};\widehat{\gamma})+(1-A_{E,i})\{1-\pi(X_{E,i};\widehat{\gamma})\}}} -{{\mu}_U(M_{E,i},X_{E,i};\lambda)\over{A_{E,i} \pi(X_{E,i};\gamma)+(1-A_{E,i})\{1-\pi(X_{E,i};\gamma)\}}}\Big] \Bigg\}.\\
\end{split}
\end{eqnarray*}
Based on (A8) and (A11), take the Taylor Expansion on above Equation at $(\gamma, \lambda)$, similar to Equation (\ref{taylor}), we have
\begin{eqnarray}\label{rearange1}
\begin{split}
\widehat{V}(\beta)-V^{\star}_n(&\beta)-\{\widehat{V}(\beta_0)-V^{\star}_n(\beta_0)\}\\
={1\over N_E}\sum_{i=1}^{N_E}\Bigg\{ \Big[\mathbb{I}&\{\phi_X(X_{E,i})^\top \beta>0\}-\mathbb{I}\{\phi_X(X_{E,i})^\top \beta_0>0\}\Big]\\
\times &\Big[-{ {\mu}_U(M_{E,i},X_{E,i};\lambda)(2A_{E,i}-1)^2\{\partial \pi (X_{E,i};\gamma)/\partial\gamma\}\over{[A_{E,i} \pi(X_{E,i};\gamma) +(1-A_{E,i})\{1-\pi(X_{E,i};\gamma)\}]^2}}(\widehat{\gamma}-{\gamma})\\
&+{(2A_{E,i}-1)\{\partial\mu_U(M_{E,i},X_{E,i};\lambda)/\partial \lambda\}\over{A_{E,i} \pi(X_{E,i};\gamma)+(1-A_{E,i})\{1-\pi(X_{E,i};\gamma)\}}}(\widehat{\lambda}-{\lambda})\Big]\Bigg\}+o_p(N_E^{-{1\over2}}).\\
\end{split}
\end{eqnarray}

Next, we define two classes of function,\\
$\mathcal{F}_\beta^{2} (x,a,m)=\Bigg\{-{ {\mu}_U(m ,x;\lambda)(2a -1)^2\{\partial \pi (x;\gamma)/\partial\gamma\} [\mathbb{I}\{\phi_X(x)^\top \beta>0\}-\mathbb{I}\{\phi_X(x)^\top \beta_0>0\} ]\over{[a \pi (x;\gamma) +(1-a )\{1-\pi(x;\gamma)\}]^2}} : ||\beta-\beta_0||_2<\delta\Bigg\}$,
and 
$
\mathcal{F}_\beta^{3} (x,a,m)=\Bigg\{{(2a-1)\{\partial\mu_U(m,x;\lambda)/\partial \lambda\} [\mathbb{I}\{\phi_X(x)^\top \beta>0\}-\mathbb{I}\{\phi_X(x)^\top \beta_0>0\} ]\over{a\pi (x;\gamma)+(1-a)\{1-\pi(x;\gamma)\}}} : ||\beta-\beta_0||_2<\delta\Bigg\}.
$

Let $M_2=\sup \big| { {\mu}_U(m ,x;\lambda)(2a -1)^2\{\partial \pi (x;\gamma)/\partial\gamma\}\over{[a \pi (x;\gamma) +(1-a )\{1-\pi(x;\gamma)\}]^2}} \big|$ and $M_3=\sup \big|{(2a-1)\{\partial\mu_U(m,x;\lambda)/\partial \lambda\}\over{a\pi (x;\gamma)+(1-a)\{1-\pi(x;\gamma)\}}}\big|$, then define the envelope of $\mathcal{F}_\beta^{j}$ as $F_j=M_j \mathbb{I}\{1-\delta\leq \phi_X(x)^\top \beta_0 \leq1+\delta\}$, for $j=2,3$. Similarly to (b2.1), we have 
\begin{eqnarray*}
||F_j||_{P,2}=M_j\sqrt{2f_X(\beta_0)} \delta^{1\over2}<\infty,
\end{eqnarray*}
and $\mathcal{F}_\beta^j$ is VC class of functions, thus, the entropy of the class function $\mathcal{F}_\beta^j$ denoted as $J_{[]}^{*}(1,\mathcal{F}^j)$ is finite, i.e., $J_{[]}^{*}(1,\mathcal{F}^j)<\infty$, for $j=2,3$. 

Then, we construct two empirical processes indexed by $\beta$,
\begin{eqnarray}\label{rearange2}
\mathbb{G}_n \mathcal{F}_\beta^j={1\over \sqrt{N_E}}\sum_{i=1}^{N_E} \Big\{\mathcal{F}_\beta^2 (X_{E,i},A_{E,i},M_{E,i})-\Mean \mathcal{F}_\beta^j (X_{E,i},A_{ i},M_{E,i})\Big\}, \quad j=2,3.
\end{eqnarray}
By Theorem 11.2 in \cite{kosorok2008introduction}, we have,
\begin{eqnarray}\label{empri}
\Mean^{*} \underset{||\beta-\beta_0||_2<\delta}{\sup} \Big|\mathbb{G}_n \mathcal{F}_\beta^j \Big|\leq c_j J_{[]}^{*}(1,\mathcal{F}^{j} ) ||F_j||_{P,2}=c_jJ_{[]}^{*}(1,\mathcal{F}^{j} )M_j\sqrt{2f_X(\beta_0)} \delta^{1\over2},\quad j=2,3,
\end{eqnarray}
where $c_2$ and $c_3$ are finite constants, and let
\begin{eqnarray*}
C_j^{*}\equiv c_jJ_{[]}^{*}(1,\mathcal{F}^{j} ) M_j\sqrt{2f_X(\beta_0)}<\infty.
\end{eqnarray*}

Finally, we rearrange the equations based on Equation (\ref{rearange1}) and Equation (\ref{rearange2}), so
\begin{eqnarray*}
\begin{split}
\eta_1= &\Mean^{*} \underset{||\beta-\beta_0||_2<\delta}{\sup} \sqrt{N_E} \Big|\widehat{V}(\beta)-V^{\star}_n(\beta)-\{\widehat{V}(\beta_0)-V^{\star}_n(\beta_0)\}\Big|\\
=&\Mean^{*} \underset{||\beta-\beta_0||_2<\delta}{\sup} \Big|(\widehat{\gamma}-{\gamma})\mathbb{G}_n \mathcal{F}_\beta^2+(\widehat{\lambda}-{\lambda})\mathbb{G}_n \mathcal{F}_\beta^3 +o_p(1)\Big|\\
\leq&{1\over \sqrt{N_E}}\Mean^{*} \underset{||\beta-\beta_0||_2<\delta}{\sup} \Big|\sqrt{N_E}(\widehat{\gamma}-{\gamma})\Big|\Big|\mathbb{G}_n \mathcal{F}_\beta^2\Big|+{1\over \sqrt{N_U}}\Mean^{*} \underset{||\beta-\beta_0||_2<\delta}{\sup} \Big|\sqrt{N_U}(\widehat{\lambda}-{\lambda})\Big|\Big|\mathbb{G}_n \mathcal{F}_\beta^3 \Big|.\\
\end{split}
\end{eqnarray*}
By the H{\"o}lder's Inequality, we have,
\begin{eqnarray*}
\begin{split}
\eta_1 \leq&{1\over \sqrt{N_E}}E \Big|\sqrt{N_E}(\widehat{\gamma}-{\gamma})\Big| \Mean^{*} \underset{||\beta-\beta_0||_2<\delta}{\sup} \Big|\mathbb{G}_n \mathcal{F}_\beta^2\Big|+{1\over \sqrt{N_U}}E \Big|\sqrt{N_U}(\widehat{\lambda}-{\lambda})\Big|\Mean^{*} \underset{||\beta-\beta_0||_2<\delta}{\sup} \Big|\mathbb{G}_n \mathcal{F}_\beta^3 \Big|.\\
\end{split}
\end{eqnarray*}
Since $\widehat{\gamma}-{\gamma}=O_E(N_E^{-{1\over2}}) $ and $\widehat{\lambda}-{\lambda}=O_E(N_U^{-{1\over2}})$, we have $M_\gamma \equiv E \Big|\sqrt{N_E}(\widehat{\gamma}-{\gamma})\Big|<\infty$ and $M_\lambda \equiv E \Big|\sqrt{N_U}(\widehat{\lambda}-{\lambda})\Big|<\infty$. By the results of (\ref{empri}) and (A9) that $t=\sqrt{{N_E\over N_U}}$ with $0<t<+\infty$, we have,
\begin{eqnarray}\label{b2}
\eta_1 \leq  {1\over \sqrt{N_E}} M_\gamma C_2^{*} \delta^{1\over2}+{1\over \sqrt{N_U}} M_\lambda C_3^{*} \delta^{1\over2} ={1\over \sqrt{N_E}} (M_\gamma C_2^{*} + tM_\lambda C_3^{*} )\delta^{1\over2}.
\end{eqnarray}

\bigskip

By the results of (\ref{b1}) in (b2.1) and (\ref{b2}) in (b2.2), we have the centered process $\widehat{V}-V$ satisfies
\begin{eqnarray*}
\begin{split}
&\Mean^{*} \underset{||\beta-\beta_0||_2<\delta}{\sup} \sqrt{N_E} |\widehat{V}(\beta)-V(\beta) -\{\widehat{V}(\beta_0)-V(\beta_0)\}|\\
\leq &\eta_1+\eta_2 \leq   C_1^{*} \delta^{1\over2}+{1\over \sqrt{N_E}} (M_\gamma C_2^{*} + tM_\lambda C_3^{*} )\delta^{1\over2}.\\
\end{split}
\end{eqnarray*}
where $M_\gamma$, $M_\lambda$, and $C_j^{*}, j=1,2,3$ are some finite constants. Let $N_E$ goes infinite, we have 
\begin{eqnarray}\label{res_lem2}
\Mean^{*} \underset{||\beta-\beta_0||_2<\delta}{\sup} \sqrt{N_E} |\widehat{V}(\beta)-V(\beta) -\{\widehat{V}(\beta_0)-V(\beta_0)\}|\leq C_1^{*} \delta^{1\over2}.
\end{eqnarray}
Let $\phi_n(\delta)= C_1^{*} \delta^{1\over2}$, and $\alpha={3\over2}<2$, check ${{\phi_n(\delta)}\over{\delta^\alpha}}={{\delta^{1\over2}}\over{\delta^{3\over2}}}=\delta^{-1}$ is decreasing not depending on $N_E$. Therefore, condition B holds.\\

(b3.) By $\widehat{\beta} \overset{p}{\longrightarrow} \beta_0$ as $N_E \to \infty$ and $
\widehat{V}(\widehat{\beta}) \geq sup_{\beta \in \rm B} \widehat{V}(\beta)$ shown previously, choose $r_n=N_E^{1/3}$, then $r_n$ satisfies
\begin{eqnarray*}
r_n^2\phi_n(r_n^{-1})=N_E^{2/3}\phi_n(N_E^{-1/3})=N_E^{2/3}(N_E^{-1/3})^{1/2}=N_E^{2/3-1/6}=N_E^{1/2}.
\end{eqnarray*}
Thus, condition C holds.\\

By the Theorem 14.4 in \cite{kosorok2008introduction}, we have $N_E^{1/3}||\widehat{\beta}-\beta_0||_2= O_p(1)$. $\square$

Note that proof of Lemma \ref{lem3} is just trivial extension of the above proof of Lemma \ref{lem3ipw}.

\subsection{Proof of Theorem \ref{thm2}}\label{ipw_proof3}

To show the asymptotical distribution of the IPW estimator of the value function, we break down the following expression $\sqrt{N_E}\big\{\widehat{V}(\widehat{\beta})-V(\beta_0)\big\}$ into two parts,
\begin{eqnarray*}
\begin{split}
&\sqrt{N_E}\big\{\widehat{V}(\widehat{\beta})-V(\beta_0)\big\}= \sqrt{N_E}\big\{\widehat{V}(\widehat{\beta})-\widehat{V}(\beta_0)+ \widehat{V}(\beta_0)-V(\beta_0)\big\}\\
=&\sqrt{N_E}\big\{\widehat{V}(\widehat{\beta})-\widehat{V}(\beta_0)\big\}+ \sqrt{N_E}\big\{\widehat{V}(\beta_0)-V(\beta_0)\big\}.\\
 \end{split}
\end{eqnarray*}

(A.) First, we show the first part 
\begin{eqnarray*}
\sqrt{N_E}\big\{\widehat{V}(\widehat{\beta})-\widehat{V}(\beta_0)\big\}=o_p(1),
\end{eqnarray*}
which is sufficient to show $\sqrt{N_E}\big\{{V}(\widehat{\beta})- {V}(\beta_0)\big\}=o_p(1)$ and $\sqrt{N_E}\Big[\{\widehat{V}( \widehat{\beta})-\widehat{V}(\beta_0)\}-\{{V}( \widehat{\beta})- {V}(\beta_0)\}\Big]=o_p(1)$.

\smallskip

(a1.) First, by $N_E^{1/3}||\widehat{\beta}-\beta_0||_2= O_p(1)$ and (A10), we take the second order Taylor expansion of $V(\widehat{\beta})$ at $\beta_0$, then
\begin{eqnarray}\label{thm2p1}
\begin{split}
&\sqrt{N_E}\{ {V}(\widehat{\beta})- {V}(\beta_0)\big\}=\sqrt{N_E}\big[ {V'(\beta_0)}||\widehat{\beta}-\beta_0||_2+{1\over 2}{V''(\beta_0)}||\widehat{\beta}-\beta_0||_2^2+o_p\{||\widehat{\beta}-\beta_0||_2^2\}\big]\\
&(\text{by }V'(\beta_0)=0)=\sqrt{N_E}\big\{ {1\over 2}{V''(\beta_0)}O_E(N_E^{-{2\over 3}})+o_p(N_E^{-{2\over 3}})\big\}= {1\over 2}{V''(\beta_0)}O_E(N_E^{-{1\over 6}})=o_p(1).\\
\end{split}
\end{eqnarray}

(a2.) Next, recall the result (\ref{res_lem2}) in the proof of Lemma \ref{lem3} that
$$
\Mean^{*} \underset{||\beta-\beta_0||_2<\delta}{\sup} \sqrt{N_E} |\widehat{V}(\beta)-V(\beta) -\{\widehat{V}(\beta_0)-V(\beta_0)\}|\leq  C_1^* \delta^{1\over2},
$$
where $C_1^*$ is a finite constant. Since $||\widehat{\beta}-\beta_0||_2 = O_p(N_E^{-1/3})$, i.e., $||\widehat{\beta}-\beta_0||_2 = c_4 N_E^{-1/3}$, where $c_4$ is a finite constant, we have,

\begin{eqnarray}\label{thm2p2}
\begin{split}
&\sqrt{N_E}\Big[\{\widehat{V}( {\widehat{\beta}})-\widehat{V}(\beta_0)\}-\{{V}( {\beta})- {V}(\beta_0)\}\Big]\\
\leq&\Mean^{*} \underset{||\beta-\beta_0||_2<c_4 N_E^{-1/3}}{\sup} \sqrt{N_E} \Big|\widehat{V}(\beta)-V(\beta) -\{\widehat{V}(\beta_0)-V(\beta_0)\}\Big|\\
\leq & C_1^* \sqrt{{c_4 N_E^{-1/3}}}= C_1^* \sqrt{c_4} N_E^{-1/6}=o_p(1).\\
\end{split}
\end{eqnarray}

(a3.) Thus, from the results of (\ref{thm2p1}) and (\ref{thm2p2}), we have,
\begin{eqnarray*}
\begin{split}
&\sqrt{N_E}\big\{\widehat{V}(\widehat{\beta})-\widehat{V}(\beta_0)\big\}\\
=& \sqrt{N_E}\Big[\{\widehat{V}( \widehat{\beta})-\widehat{V}(\beta_0)\}-\{{V}( \widehat{\beta})- {V}(\beta_0)\}\Big]+\sqrt{N_E}\big\{{V}(\widehat{\beta})- {V}(\beta_0)\big\}\\
=&o_p(1)+o_p(1)=o_p(1).
\end{split}
\end{eqnarray*}

\bigskip

(B.) Next, we only need to show the asymptotical distribution of \\
$
\sqrt{N_E}\big\{\widehat{V}(\beta_0)-V(\beta_0)\big\}= {1\over \sqrt{N_E}}\sum_{i=1}^{N_E}\Bigg[ {\mathbb{I}\{A_{E,i}=d(X_{E,i};\beta_0)\}\over{A_{E,i} \pi(X_{E,i};\widehat{\gamma})+(1-A_{E,i})\{1-\pi(X_{E,i};\widehat{\gamma})\}}} \mu_U(M_{E,i},X_{E,i};\widehat{\lambda})-V(\beta_0)\Bigg].
$

(b1.) Following the same procedure in (\ref{taylor}), by taking the Taylor Expansion on $\widehat{V}(\beta_0)$ at $( {\gamma}, {\lambda})$, we have
\begin{eqnarray}\label{tayipw}
\begin{split}
&\widehat{V}(\beta_0)=G_1^\top (\widehat{\gamma}-{\gamma})+G_2^\top (\widehat{\lambda}-{\lambda})\\
&+\underbrace{{1\over N_E}\sum_{i=1}^{N_E} {\mathbb{I}\{A_{E,i}=d(X_{E,i};\beta_0)\}\over{A_{E,i} \pi(X_{E,i}; {\gamma})+(1-A_{E,i})\{1-\pi(X_{E,i}; {\gamma})\}}} \mu_U(M_{E,i},X_{E,i}; {\lambda})}_{V^{\star}_n(\beta_0)}+o_p(N_E^{-{1\over2}}),
\end{split}
\end{eqnarray}
where $
G_1\equiv \underset{N_E\to +\infty}{\lim} {1\over N_E}\sum_{i=1}^{N_E} {\mathbb{I}\{A_{E,i}=d(X_{E,i};\beta_0)\} {\mu}_U(M_{E,i},X_{E,i};\lambda)\over{[A_{E,i} \pi(X_{E,i};\gamma) +(1-A_{E,i})\{1-\pi(X_{E,i};\gamma)\}]^2}}(1-2A_{E,i})\{\partial \pi (X_{E,i};\gamma)/\partial\gamma\}$,\\
and $
G_2\equiv \underset{N_E\to +\infty}{\lim} {1\over N_E}\sum_{i=1}^{N_E} {\mathbb{I}\{A_{E,i}=d(X_{E,i};\beta_0)\}\over{A_{E,i} \pi(X_{E,i};\gamma)+(1-A_{E,i})\{1-\pi(X_{E,i};\gamma)\}}} \{\partial\mu_U(M_{E,i},X_{E,i};\lambda)/\partial \lambda\}$.

\bigskip

(b2.) By Equation (\ref{gamma}), Equation (\ref{lambda}), and Equation (\ref{tayipw}), we have,
\begin{eqnarray*}
\begin{split}
&\sqrt{N_E}\{\widehat{V}(\beta_0)-V(\beta_0)\}\\
=&G_1^\top \sqrt{N_E}(\widehat{\gamma}-{\gamma})+G_2^\top  \sqrt{{N_E\over N_U}}\sqrt{N_U}(\widehat{\lambda}-{\lambda}) +\sqrt{N_E}\{V^{\star}_n(\beta_0)-V(\beta_0)\}+o_p(1),\\
=&\sqrt{{N_E\over N_U}} {1\over \sqrt{N_U}}  \xi_{i}^{(U)}+ {1\over \sqrt{N_E}}\sum_{i=1}^{N_E} \xi_{i}^{(E,I)}+o_p(1),\\
\end{split}
\end{eqnarray*}
where
\begin{eqnarray*}
 \xi_{i}^{(U)}\equiv G_2^\top H_2^{-1}\begin{bmatrix}
   			\phi_X(X_{U,i})\\
   			\phi_M(M_{U,i})
		\end{bmatrix} \{Y_{U,i}-{\mu}_U(M_{U,i},X_{U,i};\lambda)\}
\end{eqnarray*}
is the I.I.D. term in the auxiliary sample and 
\begin{eqnarray*}
 \xi_{i}^{(E,I)}\equiv G_1^\top  H_1^{-1} \phi_X(X_{E,i})\{A_{E,i}-\pi (X_{E,i};\gamma)\}+ {\mathbb{I}\{A_{E,i}=d(X_{E,i};\beta_0)\}\mu_U(M_{E,i},X_{E,i}; {\lambda})\over{A_{E,i} \pi(X_{E,i}; {\gamma})+(1-A_{E,i})\{1-\pi(X_{E,i}; {\gamma})\}}} -V(\beta_0)
\end{eqnarray*}
is the I.I.D. term in the experimental sample.

(b3.) By (A9), we have $t=\sqrt{{N_E\over N_U}}$ and $0<t<+\infty$. Since the estimation of $\lambda$ is independent of the experimental sample, applying the central limit theorem, we have,
\begin{eqnarray*}
\sqrt{N_E}\{\widehat{V}(\beta_0)-V(\beta_0)\}\overset{\mathcal{D}}{\longrightarrow} N(0,\sigma_{IPW}^2)
\end{eqnarray*}
where $\sigma_{IPW}^2=t\sigma_{U}^2+\sigma_{E,I}^2$, and $\sigma_{U}^2=\Mean[\{\xi_{i}^{(U)}\}^2]$ and $\sigma_{E,I}^2=\Mean[\{\xi_{i}^{(E,I)}\}^2]$.
 $\square$

\subsection{Proof of Theorem \ref{thm4}}

Proof of Theorem \ref{thm3} and Theorem \ref{thm4} are trivial extensions of the proofs of Theorem \ref{thm1} and Theorem \ref{thm2}. Here, we mainly address the augmented term of the AIPW estimator for the value function and show its asymptotic distribution.

(A.) Note that 
\begin{eqnarray*}
\begin{split}
&\widehat{V}_{AIP}(\beta) =  {1\over N_E}\sum_{i=1}^{N_E} \Big[{\mathbb{I}\{A_{E,i}=d(X_{E,i};\beta)\}\over{A_{E,i}\pi(X_{E,i};\widehat{\gamma})+(1-A_{E,i})\{1-\pi(X_{E,i};\widehat{\gamma})\}}} \mu_U(M_{E,i},X_{E,i};\widehat{\lambda})\\
&+\big\{1-{\mathbb{I}\{A_{E,i}=d(X_{E,i};\beta)\}\over{A_{E,i} \pi(X_{E,i};\widehat{\gamma})+(1-A_{E,i})\{1-\pi(X_{E,i};\widehat{\gamma})\}}} \}\widehat{E}\{\mu_U(M_{E,i},X_{E,i};\widehat{\lambda})|A_{E,i}=d(X_{E,i};\beta),X_{E,i}\}\Big].
\end{split}
\end{eqnarray*}

Following the same procedure in (\ref{taylor}), by taking the Taylor Expansion on $\widehat{V}_{AIP}(\beta) $ at $( {\gamma}, {\lambda})$, we have
\begin{eqnarray}\label{tayaipw}
\begin{split}
\widehat{V}(\beta_0)=&G_1^\top (\widehat{\gamma}-{\gamma})+G_2^\top (\widehat{\lambda}-{\lambda})\\
&+G_3^\top (\widehat{\gamma}-{\gamma})+G_4^\top (\widehat{\theta_0}-{\theta_0})+G_5^\top (\widehat{\theta_1}-{\theta_1})+V^{\star}_{n,AIP}(\beta_0)+o_p(N_E^{-{1\over2}}),
\end{split}
\end{eqnarray}
where $G_3\equiv \underset{N_E\to +\infty}{\lim} {1\over N_E}\sum_{i=1}^{N_E} - {\mathbb{I}\{A_{E,i}=d(X_{E,i};\beta_0)\} \{\phi_X(X_{E,i})^\top \theta_0+\phi_X(X_{E,i})^\top (\theta_1-\theta_0)d(X_{E,i};\beta_0)\}(1-2A_{E,i})\{\partial \pi (X_{E,i};\gamma)/\partial\gamma\}\over{[A_{E,i} \pi(X_{E,i};\gamma) +(1-A_{E,i})\{1-\pi(X_{E,i};\gamma)\}]^2}}$,\\
$G_4\equiv \underset{N_E\to +\infty}{\lim} {1\over N_E}\sum_{i=1}^{N_E}\Big[1- {\mathbb{I}\{A_{E,i}=d(X_{E,i};\beta_0)\}\over{A_{E,i} \pi(X_{E,i};\gamma)+(1-A_{E,i})\{1-\pi(X_{E,i};\gamma)\}}} \Big]\phi_X(X_{E,i})\{1-d(X_{E,i};\beta_0)\}$, \\
and $G_5\equiv \underset{N_E\to +\infty}{\lim} {1\over N_E}\sum_{i=1}^{N_E}\Big[1- {\mathbb{I}\{A_{E,i}=d(X_{E,i};\beta_0)\}\over{A_{E,i} \pi(X_{E,i};\gamma)+(1-A_{E,i})\{1-\pi(X_{E,i};\gamma)\}}} \Big]\phi_X(X_{E,i})d(X_{E,i};\beta_0)$.

\bigskip

(B.) Based on (A11) with the parametric model for $E\{\mu_U(m,x;\lambda)|A=0,X=x\}\equiv \phi_X(x)^\top \theta_0$, and $E\{\mu_U(m,x;\lambda)|A=1,X=x\}\equiv  \phi_X(x)^\top \theta_1$, we have 
\begin{eqnarray}\label{the0}
&&\sqrt{N_E}(\widehat{\theta}_0-\theta_0)\\\nonumber
&&=H_3^{-1} {1\over \sqrt{N_E}}\sum_{i=1}^{N_E} \phi_X(X_{E,i})(1-A_{E,i})\{{\mu}_U(M_{E,i},X_{E,i};\lambda)- \phi_X(X_{E,i})^\top  \theta_0\}+o_p(1),
\end{eqnarray}
where $H_3\equiv \underset{N_E\to +\infty}{\lim} {1\over N_E}\sum_{i=1}^{N_E} (1-A_{E,i})\phi_X(X_{E,i})\phi_X( X_{E,i})^\top $, and
\begin{eqnarray}\label{the1}
&&\sqrt{N_E}(\widehat{\theta}_1-\theta_1)\\\nonumber
&&=H_4^{-1} {1\over \sqrt{N_E}}\sum_{i=1}^{N_E} \phi_X(X_{E,i})A_{E,i}\{{\mu}_U(M_{E,i},X_{E,i};\lambda)- \phi_X(X_{E,i})^\top  \theta_1\}+o_p(1),
\end{eqnarray}
where $H_4\equiv \underset{N_E\to +\infty}{\lim} {1\over N_E}\sum_{i=1}^{N_E} A_{E,i}\phi_X(X_{E,i}) \phi_X(X_{E,i})^\top $.

By Equation (\ref{gamma}), Equation (\ref{lambda}), Equation (\ref{the0}), Equation (\ref{the1}), and Equation (\ref{tayaipw}), we have,
\begin{eqnarray*}
\begin{split}
&\sqrt{N_E}\big\{\widehat{V}_{AIP}(\widehat{\beta}^{G})-V(\beta_0)\big\}\\
=&G_1^\top \sqrt{N_E}(\widehat{\gamma}-{\gamma})+G_2^\top  \sqrt{{N_E\over N_U}}\sqrt{N_U}(\widehat{\lambda}-{\lambda}) +G_3^\top \sqrt{N_E}(\widehat{\gamma}-{\gamma})\\
&+G_4^\top \sqrt{N_E}(\widehat{\theta_0}-{\theta_0})+G_5^\top \sqrt{N_E}(\widehat{\theta_1}-{\theta_1})+\sqrt{N_E}\{V^{\star}_{n,AIP}(\beta_0)-V(\beta_0)\}+o_p(1),\\
=&\sqrt{{N_E\over N_U}} {1\over \sqrt{N_U}}  \xi_{i}^{(U)}+ {1\over \sqrt{N_E}}\sum_{i=1}^{N_E} \xi_{i}^{(E)}+o_p(1),\\
\end{split}
\end{eqnarray*}
where $\xi_{i}^{(E)}\equiv G_4^\top  H_3^{-1} \phi_X(X_{E,i})(1-A_{E,i})\{{\mu}_U(M_{E,i},X_{E,i};\lambda)- \phi_X(X_{E,i})^\top  \theta_0\}+\\
G_5^\top  H_4^{-1}  \phi_X(X_{E,i})A_{E,i}\{{\mu}_U(M_{E,i},X_{E,i};\lambda)- \phi_X(X_{E,i})^\top  \theta_1\}+{\mathbb{I}\{A_{E,i}=d(X_{E,i};\beta)\}{\mu}_U(M_{E,i},X_{E,i};\lambda)\over{A_{E,i}\pi(X_{E,i}; {\gamma})+(1-A_{E,i})\{1-\pi(X_{E,i}; {\gamma})\}}}+\\
\big\{1-{\mathbb{I}\{A_{E,i}=d(X_{E,i};\beta)\}\over{A_{E,i} \pi(X_{E,i}; {\gamma})+(1-A_{E,i})\{1-\pi(X_{E,i}; {\gamma})\}}} \big\}E\{{\mu}_U(M_{E,i},X_{E,i};\lambda)|A_{E,i}=d(X_{E,i};\beta),X_{E,i}\}+\\
(G_1^\top +G_3^\top ) H_1^{-1} \phi_X(X_{E,i})\{A_{E,i}-\pi (X_{E,i};\gamma)\}-V(\beta_0)$ is the I.I.D. term in the experimental sample, and $ \xi_{i}^{(U)}\equiv G_2^\top H_2^{-1}\begin{bmatrix}
   			\phi_X(X_{U,i})\\
   			\phi_M(M_{U,i})
		\end{bmatrix} \{Y_{U,i}-{\mu}_U(M_{U,i},X_{U,i};\lambda)\} $ is the I.I.D. term in the auxiliary sample.

By (A9), we have $t=\sqrt{{N_E\over N_U}}$ and $0<t<+\infty$. Since the estimation of $\lambda$ is independent of the experimental sample, applying the central limit theorem, we have,
\begin{eqnarray*}
\sqrt{N_E}\big\{\widehat{V}_{AIP}(\widehat{\beta}^{G})-V(\beta_0)\big\}\overset{\mathcal{D}}{\longrightarrow} N(0,\sigma_{AIP}^2)
\end{eqnarray*}
where $\sigma_{AIP}^2=t\sigma_{U}^2+\sigma_{E}^2$, and $\sigma_{U}^2=\Mean[\{\xi_{i}^{(U)}\}^2]$ and $\sigma_{E}^2=\Mean[\{\xi_{i}^{(E)}\}^2]$.
 $\square$

\section{Sensitivity Studies}\label{add_simu}
In this supplementary section, we investigate the finite sample performance of the proposed GEAR when the surrogacy assumption is violated in different extent, i.e. part of the information related to the long-term outcome cannot be collected or captured through intermediate outcomes. We consider the following Scenario 6 with $r=2$ and $s=2$.

\noindent \textbf{Scenario 6}:
\begin{eqnarray*}
	\left\{\begin{array}{ll}
		H^M(X)=
		\begin{bmatrix}
   			0\\
   			X^{(1)}
		\end{bmatrix},\\
		C^M(X)=
		\begin{bmatrix}
   			-0.5+0.4X^{(1)}-0.6X^{(2)}\\
   			0.5+0.6X^{(1)}-0.4X^{(2)}
		\end{bmatrix},\\
		H^Y(X)= X^{(2)},\\
		C^Y(X, M)=M^{(1)}+M^{(2)},\\
	\end{array}
	\right.
\end{eqnarray*}
where the true parameter of the ODR is $\beta_0=[0,1/\sqrt{2},-1/\sqrt{2}]^\top $ with the true value 0.333. We use the following $M^{(1)}_{par}$ as one contaminated intermediate outcome we collected instead of the original $M^{(1)}$,
\begin{eqnarray*}
M^{(1)}_{par}=M^{(1)} +A(1-l)\{-0.5+0.4X^{(1)}\},
 \end{eqnarray*}
where the parameter $l$ chosen from $\{0,0.2,0.4,0.6,0.8,1\}$ reflects the uncollected information related to the long-term outcome. When $l=1$, we have the information of intermediate outcomes is fully collected. However, under $l\in \{0,0.2,0.4,0.6,0.8\}$, the surrogacy assumption cannot hold anymore, since the long-term outcome is still dependent on the treatment given the information of $M$ and $X$.

 \begin{table} 
\centering
\caption{Empirical results of $\widehat{\beta}^G$ and its estimated value $\widehat{V}_{AIP}(\widehat{\beta}^{G})$ under the GEAR for Scenario 6 when $l=\{0,0.4,0.8\}$. Note the true parameter of the ODR is $\beta_0=[0,1/\sqrt{2},-1/\sqrt{2}]^\top $ with the true value 0.333. }\label{table:s1}
\scalebox{0.85}{
\begin{tabular}{cccc|ccc|ccc}
		\hline
		\hline
		
		&\multicolumn{3}{c|}{$l=0$} &\multicolumn{3}{c|}{$l=0.4$}&\multicolumn{3}{c}{$l=0.8$} \\
		\cline{2-10}
		& $N_E=200$ & $ 400$&$ 800$& $N_E=200$ & $400$&$800$&$N_E=200$ & $400$&$800$\\
		\hline
		\hline
 $\widehat{V}_{AIP}(\widehat{\beta}^{G})$&0.546 &0.494 & 0.470&0.505 &0.472&0.434&	0.470 &0.434 &0.401 \\
 \hline
 $SE\{\widehat{V}_{AIP}(\widehat{\beta}^{G})\}$&0.154&0.113 &0.091&0.156&0.113&0.086 &0.156 &0.118&0.088 \\
 \hline
 $\Mean\{\widehat{\sigma}_{AIP}\}$&0.158&0.118&0.092&0.160 &0.120&0.093&0.162&0.121&0.093\\
 \hline
 $V(\widehat{\beta}^G)$  &0.265&0.276&0.284  &0.285&0.296&0.298&0.293&0.306&0.314\\
 \hline
Coverage prob. (\%) &73.8 &72.8  &69.4 &82.4 &81.8 &81.8 &86.6 &86.6 &90.8 \\
 \hline
Correct Rate (\%)  &79.5  &80.2  &81.5  &83.0 &84.9 &85.2 &84.7 &87.0 &89.3 \\
 \hline
 $\ell_2$ loss of $\widehat{\beta}^G $  &0.457&0.413&0.371&0.388&0.322&0.306&0.358&0.288 &0.232\\
 \hline
$\widehat{\beta}^{(1)}$ &-0.284&-0.292&-0.276&-0.184&-0.174&-0.191&-0.041 &-0.059 &-0.054\\
 \hline
$\widehat{\beta}^{(2)}$&0.790&0.794 &0.785 &0.765&0.765&0.775&0.730&0.736 &0.739 \\
 \hline
$\widehat{\beta}^{(3)}$  &-0.544 &-0.534&-0.554&-0.618 &-0.621&-0.602 & -0.682 &-0.674&-0.671\\
 \hline
 \end{tabular}}
 \end{table}

\begin{figure}[!htp]
\centering
\includegraphics[width=5in]{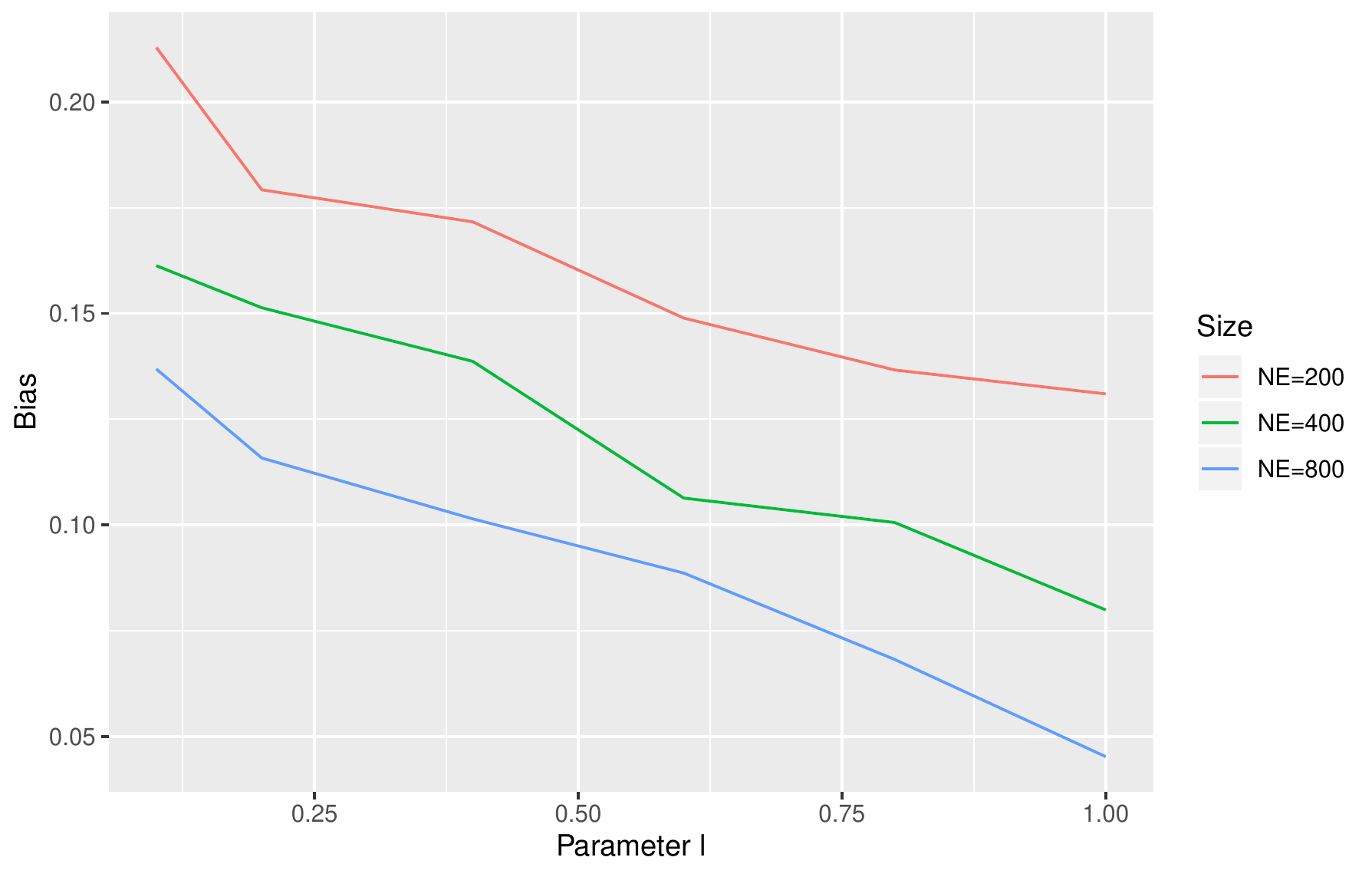}
\caption{The trend of the bias of $V(\widehat{\beta}^G)$ under the GEAR over the parameter $l$.}\label{fig: vhat1}
\end{figure}

\begin{figure}[!htp]
\centering
\includegraphics[width=5in]{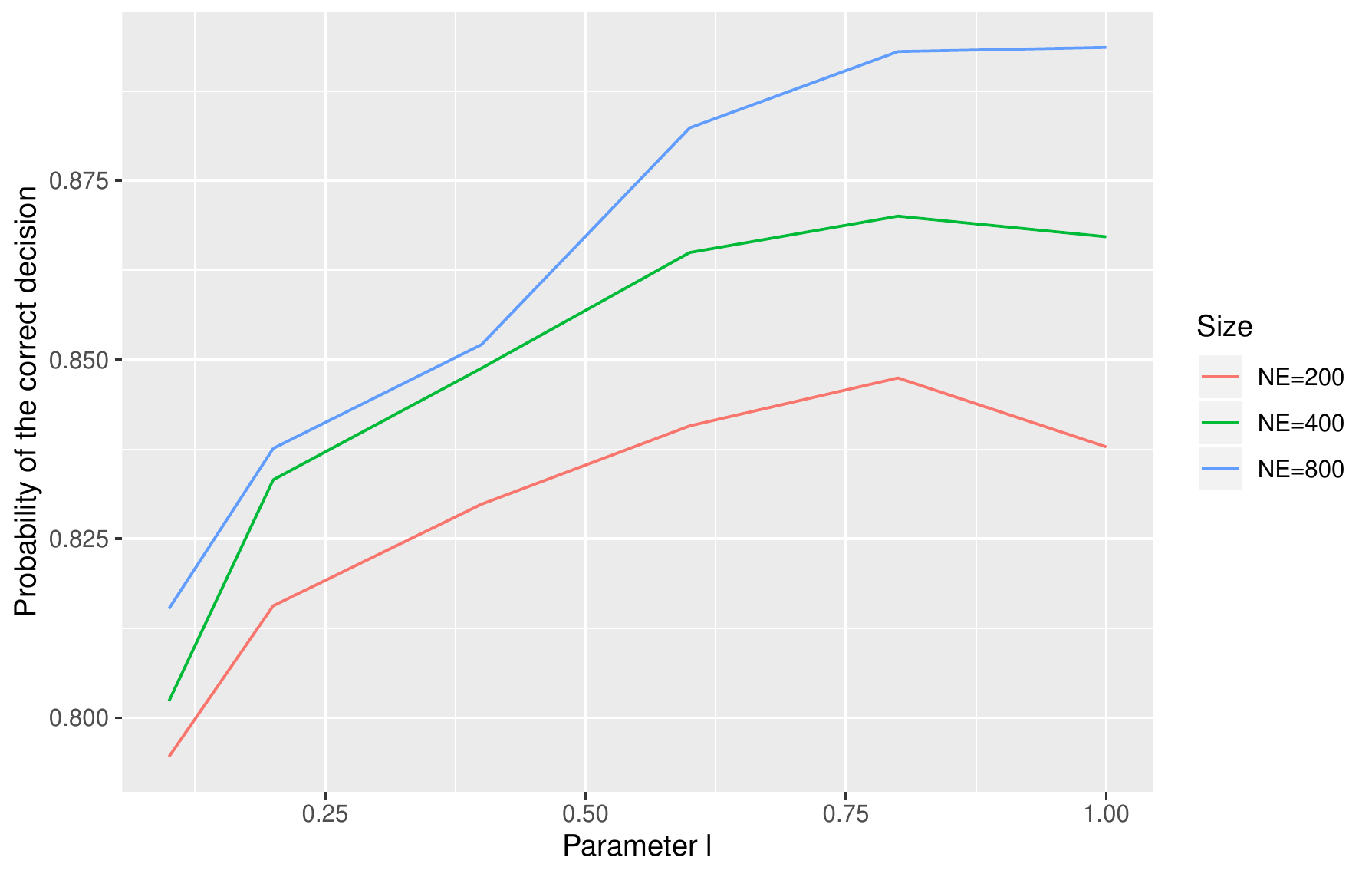}
\caption{The trend of the average rate of the correct decision made by the GEAR over the parameter $l$.}\label{fig: pcd1}
\end{figure}

Following the same estimation procedure as described in Section \ref{value_infer}, we summarize the simulation results over 500 replications in Table \ref{table:s1} for $l=\{0,0.4,0.8\}$. Figure \ref{fig: vhat1} and Figure \ref{fig: pcd1} show how the bias of $V(\widehat{\beta}^G)$ towards the true value and the average rate of the correct decision made by the GEAR change as the parameter $l$ (that indicates the uncollected information of intermediate outcomes) changes, respectively. 

Based on the results, our proposed method still has a reasonable performance when the surrogacy assumption is mildly violated. Specifically, the proposed GEAR achieves $V(\widehat{\beta}^G)=0.314$ in Scenario 6 ($V( {\beta}_0)=0.333$) with an empirical coverage probability as 90.8\% under $l=0.8$ and $N_E=800$. In addition, it is clear that including more intermediate outcomes that are highly correlated to the long-term outcome, could help to explain the treatment effect on the long-term outcome according to Figure \ref{fig: vhat1} and Figure \ref{fig: pcd1}.

\end{document}